\documentclass[12pt]{article}

\usepackage{graphicx}
\usepackage{amsmath,amssymb,amsthm}
\usepackage{amscd,fancyhdr}
\usepackage{bm,bbm}
\usepackage{graphicx, color}
\usepackage{wrapfig}

\begin{document}
\title{
\begin{flushright}
\ \\*[-80pt] 
\begin{minipage}{0.2\linewidth}
\normalsize
KIAS-P11061 \\*[50pt]
\end{minipage}
\end{flushright}
{\Large \bf Lepton Flavor Model and Decaying Dark Matter\\
 in\\
 The Binary Icosahedral Group Symmetry
\\*[20pt]}}

\author{
\centerline{
Kenji~Hashimoto$^{1}$\footnote{hashimoto@kias.re.kr} and \ Hiroshi~Okada$^{1}$\footnote{hokada@kias.re.kr}} 
\\*[20pt]
\centerline{
\begin{minipage}{\linewidth}
\begin{center}
$^1${\it \normalsize
School of Physics, KIAS, Seoul 130-722, Korea } 
\end{center}
\end{minipage}}
\\*[50pt]}
\vskip 2 cm
\date{\small
\centerline{ \bf Abstract}
\begin{minipage}{0.9\linewidth}
\medskip 
We propose a new flavor model that can simlutaneously well-describe the indirect detection experiment of (decaying) dark matter
in the cosmic-ray, with the binary icosahedral group $A'_5$.
We show not only that the $A'_{5}$ symmetry can derive quark and lepton masses and mixings consistently,
but also that are expected some predictions for the lepton sector.
And if we assume a gauge-singlet fermionic decaying dark matter,
its decay operators are also constrained by this symmetry
so that only dimension six operators of leptonic decay are allowed up to six dimensional Lagrangians.
We find that the cosmic-ray anomalies reported by PAMELA and Fermi-LAT are well explained
by decaying dark matter controlled by the $A'_{5}$ flavor symmetry, which induces universal decays. 
\end{minipage}
}

\begin{titlepage}
\maketitle
\thispagestyle{empty}
\clearpage
\thispagestyle{empty}
\tableofcontents
\thispagestyle{empty}
\end{titlepage}

\newcommand{\bi}{\bibitem}
\newcommand{\bea}{\begin{equation}}
\newcommand{\eea}{\end{equation}}
\newcommand{\be}{\begin{eqnarray}}
\newcommand{\ee}{\end{eqnarray}}
\newcommand{\nn}{\nonumber}
\newcommand{\ev}{\mbox{eV}}
\newcommand{\mev}{\mbox{MeV}}
\newcommand{\gev}{\mbox{GeV}}
\newcommand{\tev}{\mbox{TeV}}
\def\hbar#1{\backslash\hspace{-2mm}#1}
\newcommand{\OP}[1]{\operatorname{#1}}
\newcommand{\MC}[1]{\mathcal{#1}}
\newcommand{\MF}[1]{\mathfrak{#1}}
\newcommand{\Z}{\mathbb{Z}}
\newcommand{\Q}{\mathbb{Q}}
\newcommand{\R}{\mathbb{R}}
\newcommand{\C}{\mathbb{C}}
\renewcommand{\P}{\mathbb{P}}
\def\rank{\operatorname{rank}}
\newcommand{\disc}{\operatorname{disc}}
\newcommand{\ord}{\operatorname{ord}}
\newcommand{\sign}{\operatorname{sign}}
\newcommand{\CONG}{{\displaystyle\mathop{\rightarrow}^\sim}}
\newcommand{\CONGL}{\mathop{\longrightarrow}^\sim}
\newcommand{\E}{\varepsilon}
\def\2tvec#1#2{
\left(
\begin{array}{c}
#1  \\
#2  \\   
\end{array}
\right)}

\def\mat2#1#2#3#4{
\left(
\begin{array}{cc}
#1 & #2 \\
#3 & #4 \\
\end{array}
\right)
}

\def\Mat3#1#2#3#4#5#6#7#8#9{
\left(
\begin{array}{ccc}
#1 & #2 & #3 \\
#4 & #5 & #6 \\
#7 & #8 & #9 \\
\end{array}
\right)
}

\def\3tvec#1#2#3{
\left(
\begin{array}{c}
#1  \\
#2  \\   
#3  \\
\end{array}
\right)}

\def\4tvec#1#2#3#4{
\left(
\begin{array}{c}
#1  \\
#2  \\   
#3  \\
#4  \\
\end{array}
\right)}
\def\5tvec#1#2#3#4#5{
\left(
\begin{array}{c}
#1  \\
#2  \\
#3  \\
#4  \\
#5  \\
\end{array}
\right)}
\def\L{\left}
\def\R{\right}
\def\pl{\partial}
\def\lra{\leftrightarrow}
\theoremstyle{plain}
\newtheorem{thm}{Theorem}[section]
\newtheorem*{thmn}{Theorem}
\newtheorem{thmdash}[thm]{Theorem$'$}
\newtheorem*{mainthm}{Main Theorem}
\newtheorem{lem}[thm]{Lemma}
\newtheorem{cor}[thm]{Corollary}
\newtheorem{prop}[thm]{Proposition}
\newtheorem{clm}[thm]{Claim}
\newtheorem{ex}[thm]{Example}
\newtheorem{nota}[thm]{Notation}
\theoremstyle{definition}
\newtheorem{defn}[thm]{Definition}
\newtheorem{rem}[thm]{Remark}
%


\section{Introduction}

In spite of the great success of the Standard Model (SM) of the high energy physics, 
the origin of flavor structure, masses and mixings between generations, of matter particles are not uniquely determined yet. 
In order to overcome these challenges, plenty of models based on the principle of symmetry, flavor symmetry \cite{review},
have been discussed.
In particular, T2K recently reported rather large $\theta_{13}$ mixing angle of 
 the lepton mixing matrix (Maki-Nakagawa-Sakata matrix) $U_{MNS}$ \cite{Maki:1962mu}.
 It implies that one has to find how to explain the deviation from the tri-bi maximal form \cite{tribi}.
Among them, non-Abelian discrete symmetries are well discussed as plausible possibilities.

On the other hand, it has been established by the recent cosmological observations of Wilkinson Microwave Anisotropy Probe (WMAP)
that about 23 $\%$ of energy density of the universe
consists of Dark Matter (DM) \cite{Komatsu:2010fb}. And also an indirect detection experiment of dark matter (DM) in the cosmic-ray;
Fermi-LAT \cite{fermi-lat:2011rq}, recently shows the positron anomaly up to around $200$ $\gev$ energy range
as well as the total flux of $(e^++e^-)$ \cite{Abdo:2009zk,collaboration:2010ij}. It implies that 
it could be in good agreement of the PAMELA anomaly \cite{Adriani:2008zr}.
Moreover leptophilic DM is preferable, because  PAMELA measured negative results for anti-proton excess \cite{Adriani:2008zq}.
However even if the main final state of scattering or decay of DM is $\tau^+ \tau^-$,
 this annihilation/decay mode is disfavored because it overproduces gamma-rays as final state radiation \cite{Papucci:2009gd}.
This may indicate that 
these processes also reflect flavor structure of the theory when the cosmic-ray anomalies are induced by DM scattering or decay.
There are several papers in which the DM nature is related to flavor symmetry
\cite{a4-cosmic,Daikoku:2010ew,Hirsch:2010ru,Meloni:2010sk,Esteves:2010sh,Kajiyama:2006ww,Schmaltz:1994ws,Zhang:2009dd,Kajiyama:2010sb}.

In such scenarios of decaying DM, no excess of anti-proton in the cosmic-ray \cite{Adriani:2008zq} implies that 
lifetime of the DM particle should be of ${\cal O}(10^{26})$ sec. This long lifetime is achieved if 
the TeV-scale DM (gauge singlet fermion $X$) decays into leptons by dimension six operators 
$\bar L E \bar L X/\Lambda^2$ suppressed by GUT scale $\Lambda\sim 10^{16}~\gev$ \cite{decay-gut}. 
In this case, the lifetime of the DM is estimated as $\Gamma^{-1} \sim ((\tev)^5/\Lambda^4)^{-1}\sim 10^{26}$ sec. 
Remarkably, we have shown in our previous works \cite{a4-cosmic,Kajiyama:2010sb} that some Non-Abelian symmetries allowed only the term (+ mass term),
and that continuous Abelian symmetries such as $U(1)$ cannot be done in a general way \cite{a4-cosmic}.

In this paper, we show that similar argument is possible in $A'_5$ flavor symmetry model as 
an extension of the $A_4$/$T_{13}$ model. 
For the quark sector, we show the mass matrix with twelve free parameters (in the limit of no CP phases)
that can be derived by embedding three generations into 
(doublet + singlet) representations of the flavor symmetry. It is expected that nine observables can be fitted easily.
For the lepton sector, we find a promising mass matrix with only five free parameters (in the limit of no CP phases)
that can be derived by embedding three generations into triplet representations of the flavor symmetry. 
For the DM, we find the specific decay modes of leptons since the mixing matrices are already determind by our flavor symmetries.   
As a result, we find that the cosmic-ray anomalies can be well-explained by fermionic DM decay controlled by $A'_5$ symmetry, which induces universal three body decays. 

This paper is organized as follows. We briefly explain the group theory 
of $A'_5$ and show the complete list of the multiplication rules in the appendix. 
In the section 2, we construct mass matrices of the quark and lepton sector in definite choice of $A'_5$ assignment of the fields, 
and show that there exists some predictions for leptons as well as a consistent set of parameters. 
In the section 3, we show that only desirable dimension six DM decay operators are allowed by $A'_5$ symmetry 
and that leptonic decay of the DM by those operators shows good agreement with 
the cosmic-ray anomaly experiments. The section 4 is devoted to the conclusions.

\section{Matter Sector}
$A_5$ adaptability for flavor physics was initially suggested by \cite{Kajiyama:2007gx}.
Subsequently, the group was considered by these authors \cite{Everett:2008et} \cite{Chen:2010ty}
\cite{Feruglio:2011qq} \cite{Ding:2011cm}. 
The binary icosahedral group was also considered in Ref. \cite{Everett:2010rd}.
Here we find our mass matrice determined by $A'_5$ are fitted well. 
One finds the complete analysis of $A'_5$ in the appendix. Where,
as a whole notice, we define the third componet of all fields included in more than $A'_5$ triplet as negative sign is multiplied;
{\it e.g.}, $(a_1,a_2,-a_3,a_4,a_5,a_6)$.
\begin{table}[t]
\centering
\begin{tabular}{c|ccccccccc} \hline\hline
& $Q$ & $U$ & $D$ & $L$ & $E$ & $H$ &$H'$& $X$ & $\phi$ \\ \hline
$SU(2)_L \times U(1)_Y$ & 
{\bf 2}$_{1/6}$  & {\bf 1}$_{2/3}$ & {\bf 1}$_{-1/3}$ &
~{\bf 2}$_{-1/2}$~ & ~{\bf 1}$_{-1}$~  & {\bf 2}$_{1/2}$ &  ~{\bf 2}$_{1/2}$ &
~{\bf 1}$_0$ & {\bf 1}$_0$\\
$A'_{5}$ & 
${\bf (2,1)}$ & ${\bf (2,1)}$ & ${\bf  (2,1)}$ &
${\bf 3'}$ & ${\bf 3'}$ &
${\bf 3}$ &${\bf (2,1)}$ &
${\bf 1_0}$ & ${\bf 3'}$ \\ 
$Z_4$ & 
$+$ & $i$ & $-i$ & $i$ & $-$ &
$i$ &$i$ & $+$ & $-$ \\ 
\hline
\end{tabular}
\caption{\small The $A'_{5}$ and $Z_4$ charge assignment of the SM fields and the
dark matter $X$. }
\label{a'5-assign}
\end{table}

\underline{\it Quark Sector}:
We find the $A'_5\times Z_4$ invariant renormalizable Yukawa Lagrangian in the quark sector as
\be
{\cal L}^q_{Y}&=&
a_u \bar U(2) Q(1) H'(2) +b_u \bar U(1) Q(2) H'(2) +c_u \bar U(1) Q(1) H'(1)+d_u \bar U(2) Q(2) H(3)\nn \\
&+&
a_d \bar D(2) Q(1) H'^c(2) +b_d \bar D(1) Q(2) H'^c(2) +c_d \bar D(1) Q(1) H'^c(1)+d_d \bar D(2) Q(2) H^c(3)\nn\\
&+&h.c.,
\label{lag-quark}
\ee
where $H^c=\epsilon H^*$ and we assign these fields in Table \ref{a'5-assign},
and the number in parentheses of the above fields are the $A'_5$ representations.
After the electroweak symmetry breaking with vaccum alignment, {\it e.g.},
$\langle H\rangle=v$ and $\langle \phi\rangle_i=v'_i\ (i=1,2,3)$, the matrice are found as
\bea
M_u\sim
\begin{pmatrix}
d_uv   & d_uv/\sqrt2 & a_uv'_2 \cr
d_uv/\sqrt2 & d_uv & -a_uv'_1   \cr
 b_uv'_2 & -b_uv'_1  & c_uv'_3 \cr
 \end{pmatrix},\quad
M_d\sim
\begin{pmatrix}
d_dv   & d_dv/\sqrt2 & a_dv'_2 \cr
d_dv/\sqrt2 & d_dv & -a_dv'_1   \cr
 b_dv'_2 & -b_dv'_1  & c_dv'_3 \cr
 \end{pmatrix}.
\eea
As can be seen from the above matrices, there are twelve parameters (6+6) for up and down sector.
Hence one can expect that 9 observable-parameters; {\it e.g.}, 
6 quark masses and the three mixing angles in the CKM mixing matrix, are easily fitted \cite{Watanabe:2006ui}.

\underline{\it Lepton Sector}:
The charged-lepton and neutrino masses are generated 
from the $A'_{5}$ invariant operators in Table \ref{a'5-assign} as 
\bea
{\cal L}^\ell_{Y}=\frac{a_e}{\Lambda} \bar E L H^c \phi
+
\frac{a_{\nu}}{\Lambda}LHLH+h.c.,
\label{lag-lep}
\eea
where the fundamental scale is assumed to be $\Lambda={\cal O}(10^{11})~\gev$. 
After the electroweak symmetry breaking , {\it e.g.},
$\langle H\rangle=v_i$ and $\langle \phi\rangle_i=v'_i\ (i=1,2,3)$, 
the Lagrangian of Eq.({\ref{lag-lep}}) gives rise to the following mass matrix 
of charged leptons $M_e$ and neutrinos $M_{\nu}$, respectively 
\bea
M_e\sim
\frac{a_e}{\Lambda}
\begin{pmatrix}
  \sqrt2(v'_2v_1+\sqrt2 v'_1v_3) & * & * \cr
 v'_3v_2+\sqrt2 v'_1v_1 & -2v'_2v_2 & *   \cr
 v'_2v_2 &  v'_1v_2+\sqrt2 v'_3v_3 & \sqrt2(v'_2v_3+\sqrt2 v'_3v_1) \cr
 \end{pmatrix},
\eea
\bea
M_{\nu}\sim
\frac{1}{\Lambda}
\begin{pmatrix}
  \sqrt2(-a_{\nu1}+ a_{\nu2})v_1v_2 & * & * \cr
\frac{1}{\sqrt2}(-a_{\nu1}+ a_{\nu2})v^2_3 & 
2a_{\nu1}v_1v_3+a_{\nu2} \left(v_1v_3+\frac{3v^2_2}{2}\right) & * \cr
a_{\nu1}(v^2_2+v_1v_3)+a_{\nu2}  \left(2v_1v_3+\frac{v^2_2}{2}\right)&
\frac{1}{\sqrt2}(-a_{\nu1}+ a_{\nu2})v^2_1 & 
 \sqrt2(-a_{\nu1}+ a_{\nu2})v_2v_3  \cr
 \end{pmatrix}.
\eea
Notice that the symmetric matrix of $M_e$ is derived from the $A'_5$ flavor symmetry.
When vaccum alignment is achieved by $\langle H\rangle_i=v$, each of element can be simlified as
\be
M_e&\sim&
\frac{a_ev}{\Lambda}
\begin{pmatrix}
  \sqrt2(v'_2+\sqrt2 v'_1) & v'_3+\sqrt2 v'_1 & v'_2 \cr
 v'_3+\sqrt2 v'_1  & -2v'_2 & v'_1+\sqrt2 v'_3    \cr
 v'_2 &  v'_1+\sqrt2 v'_3 & \sqrt2(v'_2+\sqrt2 v'_3) \cr
 \end{pmatrix}, ~
\label{me}\\
M_{\nu}&\sim&
\frac{v^2}{\Lambda}
\begin{pmatrix}
  \sqrt2(-a_{\nu1}+ a_{\nu2}) & \frac{1}{\sqrt2}(-a_{\nu1}+ a_{\nu2}) & 2a_{\nu1}+2a_{\nu2}/5\cr
\frac{1}{\sqrt2}(-a_{\nu1}+ a_{\nu2}) & 
2a_{\nu1}+2a_{\nu2}/5 & \frac{1}{\sqrt2}(-a_{\nu1}+ a_{\nu2}) \cr
2a_{\nu1}+2a_{\nu2}/5 &
\frac{1}{\sqrt2}(-a_{\nu1}+ a_{\nu2}) & 
 \sqrt2(-a_{\nu1}+ a_{\nu2})  \cr
 \end{pmatrix}.
\label{mnu}
\ee
Here we give a numerical example to be satisfied with the exmerimental data.
When we define mixing matrices as $(m_e,m_\mu,m_\tau)=U_{eR}M_{e}U^T_{eL}$ and assume to be the real parameters, then
the mixing matrices for charged-lepton sector are found as
\be
U_{eL}=U_{eR}=\begin{pmatrix}
 -0.46193 & -0.482657 & -0.744085 \cr
 0.815049 & -0.561836 & -0.141546   \cr
 -0.349735 &  -0.67185 & 0.652919 \cr
 \end{pmatrix}, \label{uelr}
\ee
where we use $(m_e,m_\mu,m_\tau)=(0.511,105.66,1776.82)$ MeV.
For the neutrino sector, we conveniently redefine $M_\nu\rightarrow M'_\nu=O_{12}O_{23}M'_\nu O^T_{23}O^T_{12}$ first of all.
Where each of $O_{23}$ and $O_{12}$ is the reflection matrix between the flavors.
Then $M'_{\nu}$ is diagonalized by the following matrix
\be
U_{\nu}=\begin{pmatrix}
 -0.325253 & -0.350899 & 0.878112 \cr
 0.535464 & 0.699605 & 0.335269   \cr
 -0.624082 &  0.780352 & 0.0396555 \cr
 \end{pmatrix}, \label{unu1}
\ee
where we define $m_{\nu}=U_\nu M'_\nu U^T_\nu$, and use $\Delta m^2_{12}=7.65\times 10^{-5}$, $|\Delta m^2_{23}|=2.42\times 10^{-3}$
and $\sum_i m_{\nu i}\le0.01$ eV \cite{pdg10}. As a result, we find $U_{MNS}$ as
\be
|U_{MNS}|=|U^T_{eL}U_{\nu}|=\begin{pmatrix}
 0.804937 & 0.459387 & 0.0338891 \cr
 0.275433 & 0.747979 & 0.716279   \cr
 0.241252 &  0.671578 & 0.694464 \cr
 \end{pmatrix}. \label{unu2}
\ee
We are remarkably expecting that there could be three predictions in the limit of non-CP phase.
We have five free parameters, three of which come from charged-lepton sector, and two of which come from neutrino sector.
As a result, we can fit three observed mixing angles of $U_{MNS}$, 
using three charged-lepton masses and two neutrino mass differences as input paramters. 
Also they lead the normal hierarchy. However we should mention the latest result from the T2K experiment \cite{t2k} which
indicates $5.0^\circ \le \theta_{13} \le 16.0^\circ$ \cite{Fogli:2011qn} \cite{Zhang:2011aw} for the normal mass hierarchy. 
In our case  $\theta_{13} \sim 2.0^\circ$ which is smaller than the observed value. 
But it might be able to be obtained, if the vaccum alignment is relaxed. 
Finally we estimate the order of $v'$. 
To generate the tauon mass from the Eq. (\ref{me}), $vv'/\Lambda$ should be ${\cal O}$(1) $\gev$ in $a_e\simeq1$. 
Hence one finds that $v'$ be ${\cal O}(10^{8-9})$ $\gev$ because of $v={\cal O}~(100)~\gev$ and $\Lambda={\cal O}(10^{11})~\gev$.
In the next section, we will discuss the decaying DM scenario.

\section{ Decaying Dark Matter}
\begin{table}[h]
\centering
\begin{tabular}{ccl} \hline\hline
Dimensions && \multicolumn{1}{c}{DM decay operators} \\ \hline
4 && $\bar{L} H^c X$~~($H\leftrightarrow H'$) \\ 
5 && $\bar{L} H^c X\phi$ ~~($H\leftrightarrow H'$)\\
6 && 
$\bar{L}E\bar{L}X$,
~~$H^\dagger\!H\bar{L}H^cX$,
~~$(H^c)^tD_\mu H^c\bar{E}\gamma^\mu X$, \\
&&
$\bar{Q}D\bar{L}X$,
~~$\bar{U}Q\bar{L}X$,
~~$\bar{L}D\bar{Q}X$,
~~$\bar{U}\gamma_\mu D\bar{E}\gamma^\mu X$, \\
&& 
$D^\mu H^c D_\mu \bar{L} X$,
~~$D^\mu D_\mu H^c\bar{L}X$,  \\
&& 
$B_{\mu\nu}\bar{L}\sigma^{\mu\nu}H^cX$,
~~$W_{\mu\nu}^a\bar{L}\sigma^{\mu\nu}\tau^aH^cX$  \\
&&
$\bar{L} H^c X\phi^2$,
~~$H\bar L^cL\bar LX$,~~$HD^\mu H X\gamma^\mu E$
~~($H\leftrightarrow H'$)
\\ \hline
\end{tabular}
\medskip
\caption{\small The decay operators of the gauge-singlet fermionic
dark matter $X$ up to dimension six \cite{six-dim}. 
$B_{\mu\nu}$, $W_{\mu\nu}^a$, and $D_\mu$ are the field strength
tensor of hypercharge gauge boson, weak gauge boson, and the
electroweak covariant derivative.\bigskip}
\label{op}
\end{table}

 \underline{\it Decaying Dark Matter Operators}:
To well-explain the indirect detection experiments reported by PAMELA and Fermi-Lat with decaying dark matter scenario,
only the two terms are needed; 
\bea
{\cal L}^\ell_{Y}=
\frac12 m_XXX+\frac{\lambda}{\Lambda^2}\bar L E \bar L X
+h.c.,
\label{lag-dm}
\eea
where $m_X$ is the DM mass.
On the other hand, all the decay operators allowed by the SM gauge symmetry; $SU(2)_L\times U(1)_Y$, are listed in Table~\ref{op}.
Here we remarkably mention that these terms except for $\bar{L}E\bar{L}X$ are forbidden due to $A'_5\times Z_4$ symmetry
 by the field assignment of Table \ref{a'5-assign}.
Consequently the DM mainly decays into three leptons.
 With the notation
$L_i=(\nu_{i},\ell_{i})=(U_{eL})_{i \alpha}(\nu_{\alpha},\ell_{\alpha})$ 
and $E_i=(U_{eR})_{i \beta}E_{\beta}$ $(i=1,2,3,~\alpha,\beta=e,\mu,\tau)$, 
the four-Fermi decay interaction is given as 
\begin{eqnarray}
{\cal L}_{\rm decay}
&=& 
\frac{\lambda}{\Lambda^2}\,\sum_{i,j,k=1}^3(\bar{L}_iE_k)\bar{L}_kX 
\,+\text{h.c.} \nn\\
&=&\frac{\lambda}{\Lambda^2}\sum_{i,j,k=1}^3 \sum_{\alpha, \beta,\gamma=e,\mu,\tau}
\left( U_{eL}\right)_{i \alpha}^* \left( U_{eR}\right)_{j\beta}\left( U_{eL}\right)_{k \gamma}^* 
\nn\\
&\times&\left[ \left( \bar \nu_{\alpha}P_R E_{\beta}\right)\left( \bar \ell_{\gamma}P_R X\right)
-\left( \bar\ell_{\gamma}P_R E_{\beta}\right)\left( \bar \nu_{\alpha}P_R X\right)\right]
 +\text{h.c.}. \nn\\
&=&-\frac{\lambda}{\Lambda^2}
\left[ \epsilon_{\alpha\beta\gamma}\left( \bar \nu_{\alpha}P_R E_{\beta}\right)\left( \bar \ell_{\gamma}P_R X\right)
+ \epsilon_{\gamma\beta\alpha}\left( \bar\ell_{\gamma}P_R E_{\beta}\right)\left( \bar \nu_{\alpha}P_R X\right)\right]
 +\text{h.c.},\label{lag2}
\end{eqnarray}
where $\epsilon$ is an anti-symmetric tensor, which satisfies $\epsilon_{e\mu\tau}=\epsilon_{\tau e\mu}=\epsilon_{\mu\tau e}=-\epsilon_{\mu e\tau }-\epsilon_{e\tau\mu}-\epsilon_{\tau \mu e}$. 
We used the mixing matrices $U_{e(L,R)}$, which are given in Eq. ({\ref{uelr}}), from the second line to the last line in Eq. (\ref{lag2}).
One notice is as follows.
From a seven-opearator of $\bar{L} H^c X\phi^3$ that cannot be forbidden by our symmetry,
$X$ decays to $H,L$ with $(v'/\Lambda)^3$ suppresion
after the spontaneous breaking of $\phi$. Since $v'/\Lambda$ is at most ${\cal O}(10^{-3})$ to obtain the tauon mass as can be seen from the previous section,
the coupling should be ${\cal O}(10^{-12})$.

Next, we consider the branching fraction of the DM decay
through the $A'_5$ invariant Lagrangian; $\bar LE\bar L X$. Due to the
specific generation structure, $X$ decays into several tri-leptons final state with the
mixing-dependent rate. Thanks to the flavor symmetry,
the decay width of DM per each flavor 
($\Gamma_{\alpha \beta \gamma}\equiv \Gamma(X \to  \nu_{\alpha}\ell^+_{\beta}\ell^-_{\gamma})$) and
the branching fraction of each decay mode \cite{Kajiyama:2010sb} are simply given by the follwoing, respectively 
\be
&&\Gamma _{\alpha \beta \gamma}=
\frac{|\lambda|^2 m_X^5}{1536 \pi^3 \Lambda^4},
\quad
BR(X\to \nu_{\alpha}\ell^+_{\beta}\ell^-_{\gamma})=\frac{1}{6},\nn\\
&& {\rm where}\ (\alpha,\beta,\gamma)=(e,\mu ,\tau), (\tau , e,\mu),\ (\mu ,\tau , e),
 (\mu , e,\tau), (e,\tau ,\mu), (\tau , \mu , e).\label{decaywidth}
\ee
Here we have omitted the masses of charged leptons in the final states. As can be seen from the above equation, $A'_5$ symmetry induces universal three body decays.
The DM mass $m_X$ and the total decay width 
$\Gamma=\sum_{\alpha,\beta,\gamma}\Gamma_{\alpha\beta\gamma}$ 
are chosen to be free parameters in the following analysis. 
Once the decay width and the branching fractions are given, one computes the positron fraction and the total flux of positron and electron.
In this work we adopt the Navarro-Frank-White profile~\cite{Navarro:1996gj},
\begin{eqnarray}
  \rho_{\rm NFW}(\vec{x}) \,=\,
  \rho_\odot\frac{r_\odot(r_\odot+r_c)^2}{r(r+r_c)^2},
\end{eqnarray}
where $\rho_\odot\simeq0.30$~GeV/cm$^3$ is the local halo density
around the solar system, $r$ is the distance from the galactic center
whose special values $r_\odot\simeq8.5$~kpc and $r_c\simeq 20$~kpc
are the distance to the solar system and the core radius of the
profile, respectively.
We also follow Ref. \cite{a4-cosmic,Kajiyama:2010sb} for diffusion model describing the propagation of positrons and electrons 
{ \cite{Baltz:1998xv,Hooper:2004bq,Maurin:2001sj}}, and backgrounds \cite{Baltz:1998xv}. 
As can be seen from Eqs. (\ref{decaywidth}), the DM decays into various modes included in
$\tau^{\pm}$ \footnote{$\tau^{\pm}$ in the final state generally decays into hadrons, 
however such hadronic decays are suppressed by the electroweak coupling and the phase space factor.
As a result, pure leptonic decays give dominant contribution, and 
it is consistent with no anti-proton excess of the PAMELA result \cite{Adriani:2008zq}. }
 as well as $e^{\pm}$ and $\mu^{\pm}$. 
However it might be worth mentioing that the distribution function is almost the same as the $T_{13}$ model \cite{Kajiyama:2010sb},
 despite of the different decay rate. As a result, one finds in the next section that the same prediction will be found.  

\begin{figure}[t]
\begin{center}
\includegraphics[scale=0.55]{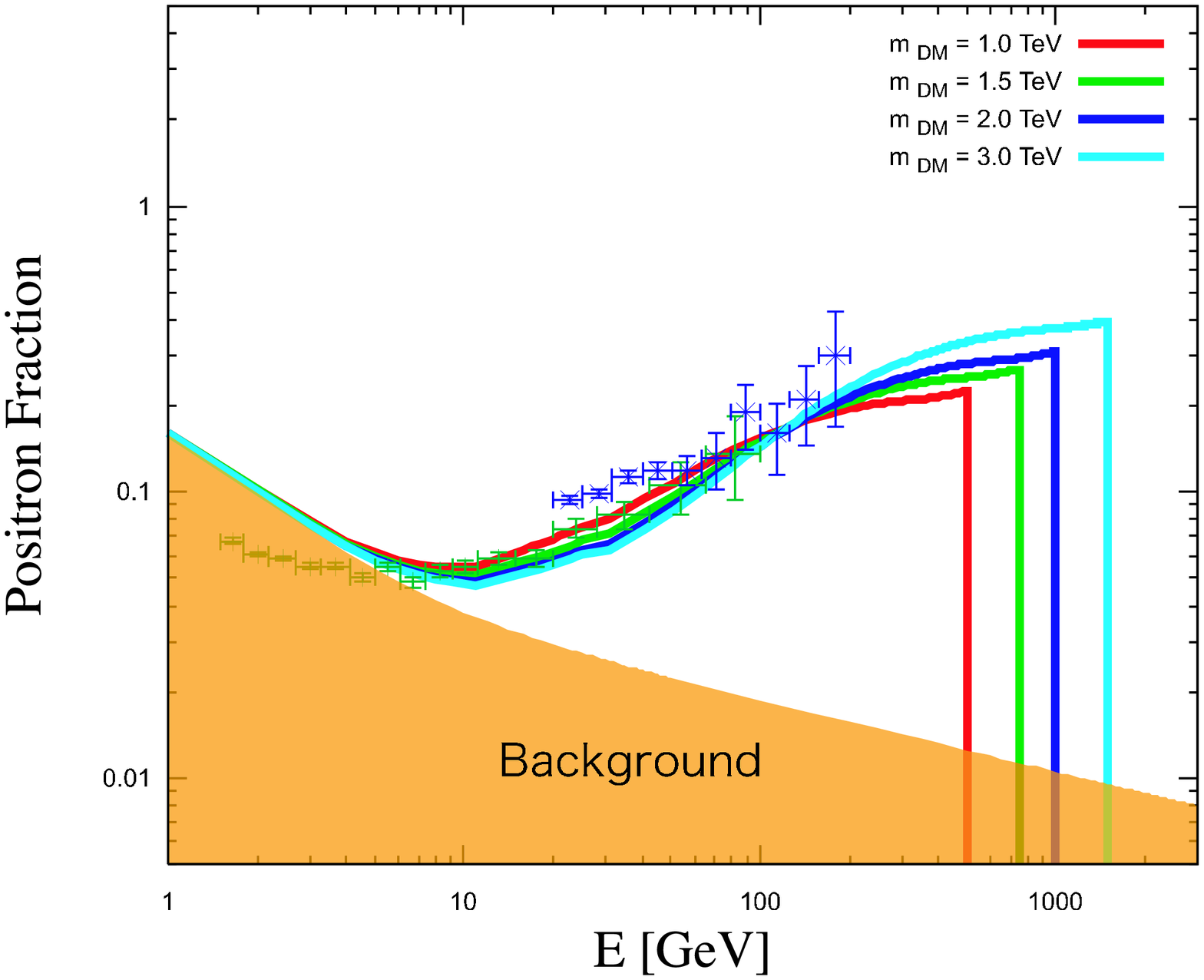}
\qquad
\includegraphics[scale=0.55]{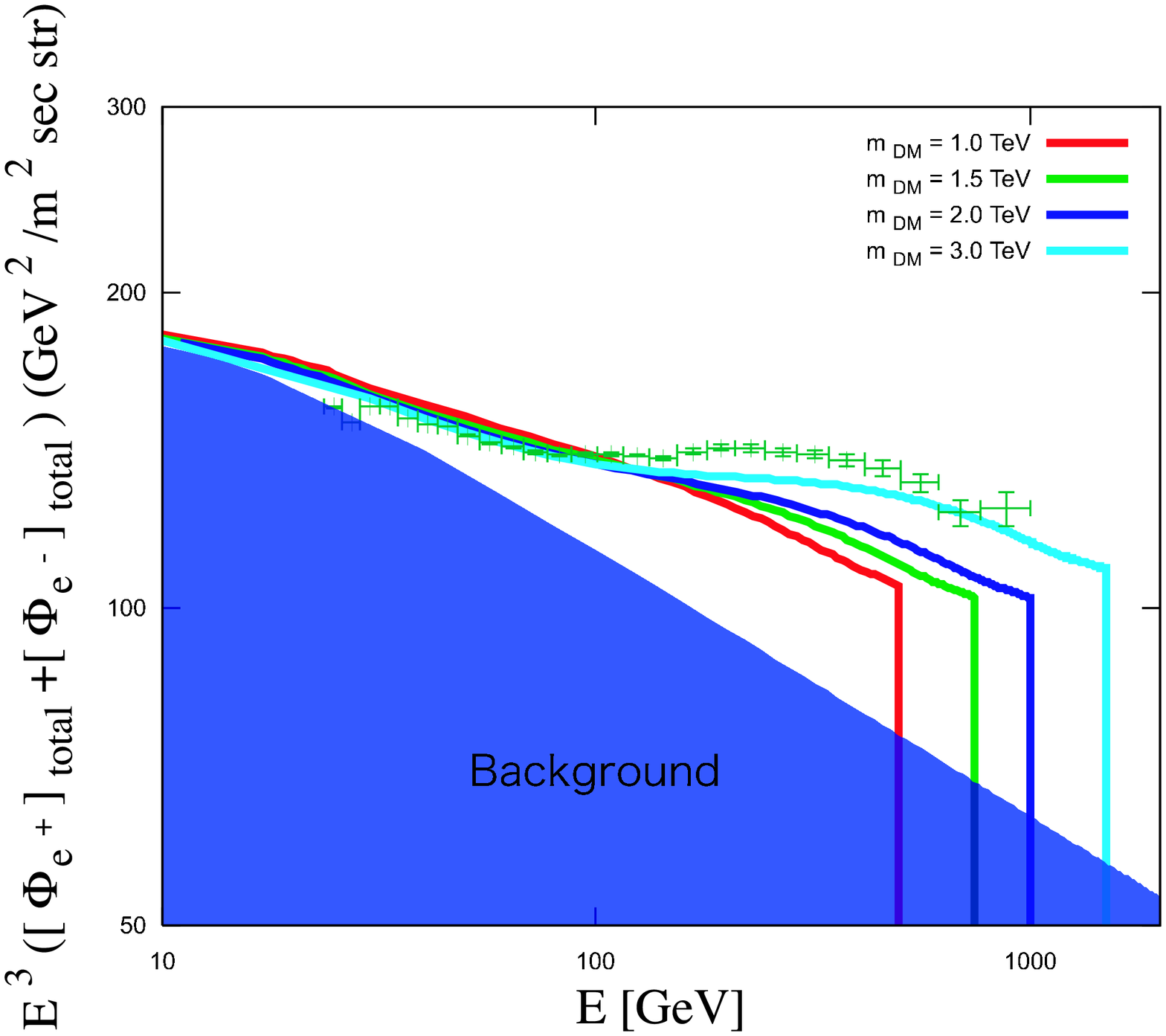}
\caption{ The lines predicted in the leptonically-decaying DM scenario with 
$A'_5$ symmetry.
 The up figure is the positron fraction, in which the green plot is PAMELA \cite{Adriani:2008zr} and
the light blue one is Fermi-LAT \cite{fermi-lat:2011rq}.
The down figure is the total $e^++e^-$ flux, in which the light green plot is Fermi-LAT \cite{Abdo:2009zk,collaboration:2010ij}.
Where the DM mass is fixed to 1.0, 1.5, 2.0, and 3.0~TeV\@. 
As for the DM decay width used in the fit, see the text.}   
\label{fig:results}
\end{center}
\end{figure}

 \underline{\it Results for PAMELA and Fermi-LAT}:
In Fig.~\ref{fig:results}, we depict the positron fraction and the total flux.
For the DM mass $m_X=1.0$, $1.5$, $2.0$, and $3.0$~TeV, we show the results
of the experimental data of PAMELA and Fermi-LAT\@. The total decay 
width $\Gamma$ is fixed for each value of DM mass so that the best fit
value explains the experimental data. With a simple $\chi^2$ analysis for the PAMELA result, we find
$\Gamma^{-1}=1.0\times10^{26}$, $8.65\times10^{25}$, $7.6\times10^{25}$, and $5.7\times10^{25}$~sec 
for $m_X=1.0$, $1.5$, $2.0$, and $3.0$~TeV, respectively. 
All the lines are well-fitted for PAMELA. 
On the other hand, the fitting lines are required rather heavy DM mass; 3 TeV $< m_X$ for Fermi-LAT. Here it might be worth mentioning that the recent gamma-ray measurement~{\cite{Dugger:2010ys,Abazajian:2010zb}}
from cluster of galaxies tells us not to have any allowed region consistent with the PAMELA and Fermi-LAT results simultaneously,
because mass and lifetime of $X$ are strongly constrained.
In order to escape these constraints, one might find that mass and lifetime
of the DM be lighter than $\sim {\cal O}(\tev)$
and longer than $\sim {\cal O}(10^{27})$ sec, respectively.
In that case, only the PAMELA results can be explained by the decaying DM.

\section{Conclusions}

We have considered a new flavor symmetric model based on a non-Abelian discrete symmetry $A'_5$. 
The form of mass matrices are determined by the assignment of $A'_5$ charges and the multiplication rules.
We have shown that masses and mixings in the quark and lepton sector are derived in the $A'_5$ model consistently. 
Furethermore, it is expected that there exist some predictions for lepton sector.

We have also shown that the decay of gauge-singlet fermionic dark matter can explain the
cosmic-ray anomalies well reported by the PAMELA and Fermi-LAT experiments, due to the universal decay coming from $A'_5$ symemtry. 
For completeness, a comprehensive analysis of mass matrices
and its phenomenology in $A'_5$ symmetric models will be published elsewhere \cite{homework}.


\section*{Acknowledgments}
H.O. would like to thank Y. Kajiyama and T. Toma for useful discussions.

\clearpage

\appendix

\section{The Binary Icosahedral Group}

We recall the binary icosahedral group $A'_5$, which is a double covering group of $A_5$. It has 120 elements and nine congugacy classes, as can be seen in Table \ref{character}. The representations of ${\bf Rep_{1}}=({\bf 1,\ 3,\ 3',\ 4,\ 5 })$ construct $A_5$ group, on the other hand the representations of ${\bf Rep_{2}}=( {\bf 2,\ 2',\ 4',\ 6 })$ come from $A'_5$ group. Then one finds
 ${\bf Rep_{1}}\otimes{\bf Rep_{1}}={\bf Rep_{1}}$,  ${\bf Rep_{1}}\otimes{\bf Rep_{2}}={\bf Rep_{2}}$, and  ${\bf Rep_{2}}\otimes{\bf Rep_{2}}={\bf Rep_{1}}$. 
 
$A'_5$ is realized as a subgroup of $SU(2)$ which is generated by
\begin{equation} \label{rep1}
 S=
 \frac{1}{\sqrt{5}}
 \begin{pmatrix}
  -(\varepsilon-\varepsilon^4)&\varepsilon^2-\varepsilon^3 \\
  \varepsilon^2-\varepsilon^3 &\varepsilon-\varepsilon^4
 \end{pmatrix}
 =
 \frac{i}{\sqrt[4]{5}}
 \begin{pmatrix} -\sqrt{\phi} & 1/\sqrt{\phi}  \\1/\sqrt{\phi} &  \sqrt{\phi}
 \end{pmatrix},\quad
 T=\begin{pmatrix}\varepsilon^3&0\\0&\varepsilon^2\end{pmatrix},
\end{equation}
 where $\E=\exp(2 \pi i/5)$ and
 $\phi=-(\E^2+\E^3)=(1+\sqrt{5})/2$
 (see \cite{klein}).
In fact, $A'_5$ has a presentation
\begin{equation}
 \langle s,t \bigm| s^2=(st)^3=t^5 \rangle
\end{equation}
 and $S,T$ satisfy
\begin{equation}
 S^2=(-ST)^3=(-T)^5=-I.
\end{equation}
We have another realization of $A'_5$ as follows:
\begin{equation} \label{rep2}
 S'=\frac{-1}{\sqrt{5}}
 \begin{pmatrix}
  -(\varepsilon^2-\varepsilon^3)&\varepsilon^4-\varepsilon \\
  \varepsilon^4-\varepsilon &\varepsilon^2-\varepsilon^3
 \end{pmatrix}
 =\frac{i}{\sqrt[4]{5}}
 \begin{pmatrix}
  1/\sqrt{\phi} & \sqrt{\phi} \\
  \sqrt{\phi} & -1/\sqrt{\phi}
 \end{pmatrix},
 T'=\begin{pmatrix}\varepsilon&0\\0&\varepsilon^4\end{pmatrix}.
\end{equation}

\begin{table}[tbp]
\begin{tabular}{|c|c|c|c|c|c||c|c|c|c|}
\hline
$\mathcal{I^{\prime}}$& $\;$$\;$ \textbf{1} $\;$$\;$&\textbf{3}&$\textbf{3}^{\prime}$& $\;$$\;$ \textbf{4} $\;$$\;$ & $\;$$\;$ \textbf{5} $\;$$\;$ &\textbf{2}&\textbf{2}$^{\prime}$& $\;$$\;$ \textbf{4}$^{\prime}$ $\;$$\;$&$\;$ $\;$ \textbf{6}$\;$ $\;$ \\ \hline
\textbf{$1$}& 1 &3&3&4&5&2&2&4&6\\ \hline
\textbf{$12 C_5$}&1&$\phi$&$1-\phi$&$-$1&0&$-\phi$&$\phi-1$&$-1$&1\\ \hline
$12{C}^2_{5}$&1&$1-\phi$&$\phi$&$-$1&0&$\phi -1$&$-\phi$&$-$1&1\\ \hline
$20{C}_{3}$&1&0&0&1&$-$1&$-1$&$-1$&$1$&0\\ \hline
$30 {C}_{2}$&1&$-$1&$-$1&0&1&0&0&0&0 \\ \hline
\hline
$R$&1&3&3&4&5&$-$2&$-$2&$-$4&$-$6\\ \hline
$12 C_5R$&1&$\phi$&$1-\phi$&$-$1&0&$\phi$&$1-\phi$&1&$-1$\\ \hline
$12 C_5^2R$&1&$1-\phi$&$\phi$&$-$1&0&$1-\phi$&$\phi$&1&$-1$\\ \hline
$20C_3R$&1&0&0&1&$-$1&1&1&$-1$&0\\ \hline
\end{tabular}
\caption{Character table of $A'_5$ }
\label{character}
\end{table}

\section{Representations of $A'_5$}

Let $\rho_1=(x_1,x_2),\rho'_1=(x'_1,x'_2)$ be
 the fundamental representations
 determined by Eqs.\ (\ref{rep1}) and (\ref{rep2}), respectively.
We define $\rho_n$($\rho'_n$) as
 the $n$-th symmetric products of $\rho_1$($\rho'_1$),
 which are written by
\begin{gather}
 \rho_n=(
  x^n_1,
  \sqrt{\begin{pmatrix}n\\1\end{pmatrix}} x^{n-1}_1 x_2,
  \sqrt{\begin{pmatrix}n\\2\end{pmatrix}} x^{n-2}_1 x^2_2,
  \ldots,
  x^n_2
 ), \\
 \rho'_n=(
  (x'_1)^n,
  \sqrt{\begin{pmatrix}n\\1\end{pmatrix}} (x'_1)^{n-1} x'_2,
  \sqrt{\begin{pmatrix}n\\2\end{pmatrix}} (x'_1)^{n-2} (x'_2)^2,
  \ldots,
  (x'_2)^n
 ),
\end{gather}
 respectively.
Here $(\begin{smallmatrix}n\\i\end{smallmatrix})$
 is a binomial coefficient.
We can check
\begin{gather}
 \rho_1\sim{\bf 2},\rho_2\sim{\bf 3},\rho_3\sim{\bf 4'},
 \rho_4\sim{\bf 5},\rho_5\sim{\bf 6}, \\
 \rho_6\sim{\bf 3'}\oplus{\bf 4},
 \rho_7\sim{\bf 2'}\oplus{\bf 6},
 \rho_8\sim{\bf 4}\oplus{\bf 5},
 \rho_9\sim{\bf 4'}\oplus{\bf 6},
 \rho_{10}\sim{\bf 3}\oplus{\bf 3'}\oplus{\bf 5}
\end{gather}
 by direct computation.
Since $\rho'_n$ is ``conjugate'' to $\rho_n$,
 we have $\rho'_1\sim{\bf 2'},\rho'_2\sim{\bf 3'}$.
We define representation matrices of irreducible representations
 of $A'_5$ as follows.

${\bf 2}=\rho_1$
\begin{equation}
 S_2=
 \frac{i}{\sqrt[4]{5}}
 \begin{pmatrix} -\sqrt{\phi} & 1/\sqrt{\phi}  \\1/\sqrt{\phi} &  \sqrt{\phi}
 \end{pmatrix},\quad
 T_2=\begin{pmatrix}\varepsilon^3&0\\0&\varepsilon^2\end{pmatrix},
\end{equation}

${\bf 2'}=\rho'_1$
\begin{equation}
 S_{2'}=
  \frac{i}{\sqrt[4]{5}}
 \begin{pmatrix}
  1/\sqrt{\phi} & \sqrt{\phi} \\
\sqrt{\phi} & -1/\sqrt{\phi}
 \end{pmatrix},
 T_{2'}=\begin{pmatrix}\varepsilon&0\\0&\varepsilon^4\end{pmatrix},
\end{equation}

${\bf 3}=\rho_2$
\begin{equation}
S_3= \frac{-1}{\sqrt{5}}
 \begin{pmatrix}
  \phi&-\sqrt{2}&1/\phi\cr
   -\sqrt{2}&-1&  \sqrt{2}\cr
    1/\phi&\sqrt{2}&\phi\cr
 \end{pmatrix},\quad
T_3= \begin{pmatrix}
  \E&0&0\cr 0&1&0\cr 0&0&\E^4\cr
 \end{pmatrix},
\end{equation}

${\bf 3'}=\rho'_2$
\begin{equation}
S_{3'}= \frac{1}{\sqrt{5}}
 \begin{pmatrix}
  -1/\phi&-\sqrt{2}& -\phi\cr 
  -\sqrt{2}&-1&  \sqrt{2}\cr 
  -\phi&\sqrt{2}&-1/\phi\cr
 \end{pmatrix},\quad
T_{3'}= \begin{pmatrix}
  \E^2&0&0\cr 0&1&0\cr 0&0&\E^3\cr
 \end{pmatrix},
\end{equation}

${\bf 4}={\bf 2}\otimes{\bf 2'}=\rho_1\otimes\rho'_1$
\begin{equation}
 S_4=
 \frac{-1}{\sqrt{5}}
 \begin{pmatrix}
   -1  & -\phi & 1/\phi & 1  \\
  -\phi&     1 &      1 & -1/\phi \\
   1/\phi & 1       & 1    & \phi \\
        1 & -1/\phi & \phi & -1 
 \end{pmatrix},\quad
 T_4=\begin{pmatrix}
  \E^4 \\
 & \E^2 \\
 && \E^3 \\
 &&& \E
 \end{pmatrix},
\end{equation}

${\bf 4'}=\rho_3$
\begin{equation}
S_{4'}= \frac{-i}{5^{3/4} \sqrt{\phi}}
 \begin{pmatrix}
  -\left(\phi+1\right)&\sqrt{3}\,\phi&-\sqrt{3}&\phi-1\cr
  \sqrt{3}\,\phi&\phi-1&-\left(\phi+1\right)&\sqrt{3}\cr
  -\sqrt{3}&-\left(\phi+1
  \right)&-\left(\phi-1\right)&\sqrt{3}\,\phi\cr
  \phi-1&\sqrt{3}&\sqrt{3}\,\phi&\phi+1\cr
 \end{pmatrix},\quad
T_{4'}= \begin{pmatrix}
  \E^4&0&0&0\cr 0&\E^3&0&0\cr 0&0&\E^2&0
 \cr 0&0&0&\E\cr
 \end{pmatrix},
\end{equation}

${\bf 5}=\rho_4$
\begin{equation}
S_5= \frac{1}{5}
 \begin{pmatrix}
  \phi+1&-2\,\phi&\sqrt{6}& 2\left(1-\phi\right)& \left(2-\phi\right) \cr 
 -2\,\phi& \left(2-\phi\right)&\sqrt{6}&-\left(\phi+1\right)&2 \left(\phi-1\right)\cr 
 \sqrt{6}&\sqrt{6}&-1&-\sqrt{6}&\sqrt{6} \cr
  2\left(1-\phi\right)&-\left(\phi+1\right)&-
 \sqrt{6}&-\left(\phi-2\right)&2\,\phi\cr 
 \left(2-\phi\right)&2\,\left(\phi-1\right)&\sqrt{6}&2\,\phi& \phi+1\cr
 \end{pmatrix},\
T_5=
 \begin{pmatrix}
  \E^{2}&0&0&0&0\cr 0&\E&0&0&0\cr 0&0&
  1&0&0\cr 0&0&0&\E^4&0\cr 0&0&0&0&\E^3\cr
 \end{pmatrix},
\end{equation}

${\bf 6}=\rho_5$
\begin{gather}
S_6= \frac{i}{5^{5/4} \sqrt{\phi}}\times \\
{\small
 \begin{pmatrix}
  -\left(2\,\phi+1\right)&\sqrt{5}\left(\phi+1\right)
 &-\sqrt{10}\phi&
 \sqrt{10}&-\sqrt{5}\left(\phi-1\right)&-\left(
 \phi-2\right)\cr
 \sqrt{5}\,\left(\phi+1\right)&-\left(2\,
 \phi-1\right)&-{{\sqrt{2}\,\left(2\,\phi-1\right)}}
 &{{\sqrt{2}\,\left(\phi+2\right)}}
 &-\left(\phi+2\right)&\sqrt{5}\,\left(\phi-1\right)\cr
 -\sqrt{10}\,\phi&-{{\sqrt{2}\,\left(2\,\phi-1\right)}}
 &\phi+2&2\,\phi-1&-{{\sqrt{2}\,\left(\phi+2
 \right)}}&\sqrt{10}\cr
 \sqrt{10}&{{\sqrt{2}\,\left(
 \phi+2\right)}}&2\,\phi-1&-\left(\phi+2
 \right)&-{{\sqrt{2}\,\left(2\,\phi-1\right)}}&
 \sqrt{10}\,\phi\cr
 -\sqrt{5}\,\left(\phi-1\right)&-\left(
 \phi+2\right)&-{{\sqrt{2}\,\left(\phi+2\right)}}
 &-{{\sqrt{2}\,\left(2\,\phi-1\right)}}
 &2\,\phi-1&\sqrt{5}\,\left(\phi+1\right)\cr
 -\left(
 \phi-2\right)&{{\sqrt{5}\,\left(\phi-1\right)}}&{{\sqrt{10}
 }}&{{\sqrt{10}\,\phi}}&{{\sqrt{5}\,\left(
 \phi+1\right)}}&2\,\phi+1\cr
 \end{pmatrix}
},
\end{gather}
\begin{equation}
T_6= \begin{pmatrix}
  1&0&0&0&0&0\cr 0&\E^{4}&0&0&0&0\cr 0&0&
 \E^{3}&0&0&0\cr 0&0&0&\E^{2}&0&0\cr 0&0&0&0&\E
 &0\cr 0&0&0&0&0&1\cr
 \end{pmatrix}.
\end{equation}

\section{Tensor Products}
Now we give the list of decompositions
 of tensor products of irreducible representations of $A'_5$
 under our bases.
Products of characters of $A'_5$
 are decomposed as shown in Table \ref{decomposition}.
To decompose $\rho_n\otimes\rho_m$ ($n \leq m$), we apply
 Clebsch--Gordan formula
\begin{align}
 & (x_1^n,\ldots,x_2^n)_{\rho_n} \otimes
 (y_1^m,\ldots,y_2^m)_{\rho'_m} \notag \\
 =&
 (y_1^{m-n} (x_1 y_2-x_2 y_1)^{n},\ldots )_{\rho_{m-n}}
 \oplus
 (x_1 y_1^{m-n+1} (x_1 y_2-x_2 y_1)^{n-1},\ldots )_{\rho_{m-n+2}}
 \oplus \cdots \\ & \oplus
 (x_1^{n} y_1^{m},\ldots )_{\rho_{n+m}}. \notag
\end{align}
For example, we have
\begin{equation}
 {\bf 4} \otimes {\bf 5'} =\rho_3\otimes \rho_4
 = \rho_1 \oplus \rho_3 \oplus \rho_5 \oplus \rho_7
 = {\bf (2)\oplus (4')\oplus (6) \oplus (2'\oplus 6) }.
\end{equation}

\begin{table}[t]
\centering
\begin{equation*}
\begin{array}{c|ccccccccc}
&{\bf1}&{\bf2}&{\bf 2'}&\bf3&\bf3'&\bf4&\bf4'&\bf5&\bf6 \\
\hline
\bf1&\bf1&\bf2&\bf2'&\bf3&\bf3'&\bf4&\bf4'&\bf5&\bf6 \\
\bf2&&\bf13&\bf4&\bf24'&\bf6&\bf2'6&\bf35&\bf4'6&\bf3'45 \\
\bf2'&&&\bf13'&\bf6&\bf2'4'&\bf26&\bf3'5&\bf4'6&\bf345 \\
\bf3&&&&\bf135&\bf45&\bf3'45&\bf24'6&\bf33'45&\bf2'4'66 \\
\bf3'&&&&&\bf13'5&\bf345&\bf2'4'6&\bf33'45&\bf24'66 \\
\bf4&&&&&&\bf133'45&\bf4'66&\bf33'455&\bf22'4'4'66 \\
\bf4'&&&&&&&\bf133'45&\bf22'4'66&\bf33'4455 \\
\bf5&&&&&&&&\bf133'4455&\bf22'4'4'666 \\
\bf6&&&&&&&&&\bf1333'3'44555
\\\hline
\end{array}
\end{equation*}
\caption{Decomposition rules of $A'_5$ }
\label{decomposition}
\end{table}

We write down the whole tensor products below.

$\bf2\otimes2=1\oplus3 $
$$ {\bf 2} \otimes {\bf 2} : {\bf 1} $$ 
$$\left( { x_1}\,{ y_2}-{ x_2}\,{ y_1} \right) $$
$$ {\bf 2} \otimes {\bf 2} : {\bf 3} $$ 
$$\left( { x_1}\,{ y_1} , {{{ x_1}\,{ y_2}}\over{\sqrt{2}
 }}+{{{ x_2}\,{ y_1}}\over{\sqrt{2}}} , { x_2}\,{ y_2}
  \right) $$
  
$\bf2\otimes2'=4$ 
$$ {\bf 2} \otimes {\bf 2'} : {\bf 4} $$ 
$$\left( { x_1}\,{ y_1} , { x_1}\,{ y_2} , { x_2}\,
 { y_1} , { x_2}\,{ y_2} \right) $$
 
$\bf2\otimes3=2\oplus4'$ 
$$ {\bf 2} \otimes {\bf 3} : {\bf 2} $$ 
$$\left( {{{ x_1}\,{ y_2}}\over{\sqrt{3}}}-{{2\,{ x_2}\,
 { y_1}}\over{\sqrt{6}}} , {{2\,{ x_1}\,{ y_3}}\over{\sqrt{6
 }}}-{{{ x_2}\,{ y_2}}\over{\sqrt{3}}} \right) $$
$$ {\bf 2} \otimes {\bf 3} : {\bf 4'} $$ 
$$\left( { x_1}\,{ y_1} , {{2\,{ x_1}\,{ y_2}}\over{
 \sqrt{6}}}+{{{ x_2}\,{ y_1}}\over{\sqrt{3}}} , {{{ x_1}\,
 { y_3}}\over{\sqrt{3}}}+{{2\,{ x_2}\,{ y_2}}\over{\sqrt{6}
 }} , { x_2}\,{ y_3} \right) $$
 
$\bf2\otimes3'=6$ 
$$ {\bf 2} \otimes {\bf 3'} : {\bf 6} $$ 
$$\left( -{{2\,{ x_2}\,{ y_3}}\over{\sqrt{5}}}-{{{ x_1}\,
 { y_1}}\over{\sqrt{5}}} , { x_2}\,{ y_1} , { x_1}\,
 { y_2} , -{ x_2}\,{ y_2} , { x_1}\,{ y_3} , {{2\,
 { x_1}\,{ y_1}}\over{\sqrt{5}}}-{{{ x_2}\,{ y_3}}\over{
 \sqrt{5}}} \right) $$
 
$\bf2\otimes4=2'\oplus6$ 
$$ {\bf 2} \otimes {\bf 4} : {\bf 2'} $$ 
$$\left( {{{ x_1}\,{ y_3}}\over{\sqrt{2}}}-{{{ x_2}\,
 { y_1}}\over{\sqrt{2}}} , {{{ x_1}\,{ y_4}}\over{\sqrt{2}}}
 -{{{ x_2}\,{ y_2}}\over{\sqrt{2}}} \right) $$
$$ {\bf 2} \otimes {\bf 4} : {\bf 6} $$ 
$$\left( {{3\,{ x_2}\,{ y_3}}\over{\sqrt{10}}}+{{{ x_1}\,
 { y_2}}\over{\sqrt{10}}} , -{{{ x_1}\,{ y_4}}\over{\sqrt{2}
 }}-{{{ x_2}\,{ y_2}}\over{\sqrt{2}}} , { x_2}\,{ y_4} , 
 -{ x_1}\,{ y_1} , {{{ x_1}\,{ y_3}}\over{\sqrt{2}}}+{{
 { x_2}\,{ y_1}}\over{\sqrt{2}}} , {{3\,{ x_1}\,{ y_2}
 }\over{\sqrt{10}}}-{{{ x_2}\,{ y_3}}\over{\sqrt{10}}} \right) $$

$\bf2\otimes4'=3\oplus5$ 
$$ {\bf 2} \otimes {\bf 4'} : {\bf 3} $$ 
$$\left( {{{ x_1}\,{ y_2}}\over{2}}-{{3\,{ x_2}\,{ y_1}
 }\over{2\,\sqrt{3}}} , {{{ x_1}\,{ y_3}}\over{\sqrt{2}}}-{{
 { x_2}\,{ y_2}}\over{\sqrt{2}}} , {{3\,{ x_1}\,{ y_4}
 }\over{2\,\sqrt{3}}}-{{{ x_2}\,{ y_3}}\over{2}} \right) $$
$$ {\bf 2} \otimes {\bf 4'} : {\bf 5} $$ 
$$\left( { x_1}\,{ y_1} , {{3\,{ x_1}\,{ y_2}}\over{2\,
 \sqrt{3}}}+{{{ x_2}\,{ y_1}}\over{2}} , {{{ x_1}\,{ y_3}
 }\over{\sqrt{2}}}+{{{ x_2}\,{ y_2}}\over{\sqrt{2}}} , {{
 { x_1}\,{ y_4}}\over{2}}+{{3\,{ x_2}\,{ y_3}}\over{2\,
 \sqrt{3}}} , { x_2}\,{ y_4} \right) $$
 
$\bf2\otimes5=4'\oplus6$ 
$$ {\bf 2} \otimes {\bf 5} : {\bf 4'} $$ 
$$\left( {{{ x_1}\,{ y_2}}\over{\sqrt{5}}}-{{2\,{ x_2}\,
 { y_1}}\over{\sqrt{5}}} , {{2\,{ x_1}\,{ y_3}}\over{\sqrt{10
 }}}-{{3\,{ x_2}\,{ y_2}}\over{\sqrt{15}}} , {{3\,{ x_1}\,
 { y_4}}\over{\sqrt{15}}}-{{2\,{ x_2}\,{ y_3}}\over{\sqrt{10
 }}} , {{2\,{ x_1}\,{ y_5}}\over{\sqrt{5}}}-{{{ x_2}\,
 { y_4}}\over{\sqrt{5}}} \right) $$
$$ {\bf 2} \otimes {\bf 5} : {\bf 6} $$ 
$$\left( { x_1}\,{ y_1} , {{2\,{ x_1}\,{ y_2}}\over{
 \sqrt{5}}}+{{{ x_2}\,{ y_1}}\over{\sqrt{5}}} , {{3\,{ x_1}
 \,{ y_3}}\over{\sqrt{15}}}+{{2\,{ x_2}\,{ y_2}}\over{\sqrt{10
 }}} , {{2\,{ x_1}\,{ y_4}}\over{\sqrt{10}}}+{{3\,{ x_2}\,
 { y_3}}\over{\sqrt{15}}} , {{{ x_1}\,{ y_5}}\over{\sqrt{5}
 }}+{{2\,{ x_2}\,{ y_4}}\over{\sqrt{5}}} , { x_2}\,{ y_5}
  \right) $$
  
$\bf2\otimes6=3'\oplus4\oplus5$ 
$$ {\bf 2} \otimes {\bf 6} : {\bf 3'} $$ 
$$\left( -{{2\,{ x_2}\,{ y_6}}\over{\sqrt{10}}}+{{{ x_1}\,
 { y_2}}\over{\sqrt{2}}}+{{{ x_2}\,{ y_1}}\over{\sqrt{10}}}
  , -{{{ x_1}\,{ y_4}}\over{\sqrt{2}}}-{{{ x_2}\,{ y_3}
 }\over{\sqrt{2}}} , -{{{ x_1}\,{ y_6}}\over{\sqrt{10}}}-{{
 { x_2}\,{ y_5}}\over{\sqrt{2}}}-{{2\,{ x_1}\,{ y_1}
 }\over{\sqrt{10}}} \right) $$
$$ {\bf 2} \otimes {\bf 6} : {\bf 4} $$ 
$$\left( {{{ x_1}\,{ y_5}}\over{\sqrt{3}}}+{{2\,{ x_2}\,
 { y_4}}\over{\sqrt{6}}} , -{{3\,{ x_2}\,{ y_6}}\over{\sqrt{15
 }}}-{{{ x_1}\,{ y_2}}\over{\sqrt{3}}}-{{{ x_2}\,{ y_1}
 }\over{\sqrt{15}}} , -{{{ x_1}\,{ y_6}}\over{\sqrt{15}}}-{{
 { x_2}\,{ y_5}}\over{\sqrt{3}}}+{{3\,{ x_1}\,{ y_1}
 }\over{\sqrt{15}}} , {{2\,{ x_1}\,{ y_3}}\over{\sqrt{6}}}+{{
 { x_2}\,{ y_2}}\over{\sqrt{3}}} \right) $$
$$ {\bf 2} \otimes {\bf 6} : {\bf 5} $$ 
$$\left( {{{ x_1}\,{ y_2}}\over{\sqrt{6}}}-{{5\,{ x_2}\,
 { y_1}}\over{\sqrt{30}}} , {{{ x_1}\,{ y_3}}\over{\sqrt{3}
 }}-{{2\,{ x_2}\,{ y_2}}\over{\sqrt{6}}} , {{{ x_1}\,
 { y_4}}\over{\sqrt{2}}}-{{{ x_2}\,{ y_3}}\over{\sqrt{2}}}
  , {{2\,{ x_1}\,{ y_5}}\over{\sqrt{6}}}-{{{ x_2}\,{ y_4}
 }\over{\sqrt{3}}} , {{5\,{ x_1}\,{ y_6}}\over{\sqrt{30}}}-{{
 { x_2}\,{ y_5}}\over{\sqrt{6}}} \right) $$
 
$\bf2'\otimes2'=1\oplus3'$ 
$$ {\bf 2'} \otimes {\bf 2'} : {\bf 1} $$ 
$$\left( { x_1}\,{ y_2}-{ x_2}\,{ y_1} \right) $$
$$ {\bf 2'} \otimes {\bf 2'} : {\bf 3'} $$ 
$$\left( { x_1}\,{ y_1} , {{{ x_1}\,{ y_2}}\over{\sqrt{2}
 }}+{{{ x_2}\,{ y_1}}\over{\sqrt{2}}} , { x_2}\,{ y_2}
  \right) $$
  
$\bf2'\otimes3=6$ 
$$ {\bf 2'} \otimes {\bf 3} : {\bf 6} $$ 
$$\left( {{3\,{ x_1}\,{ y_3}}\over{\sqrt{10}}}+{{{ x_2}\,
 { y_1}}\over{\sqrt{10}}} , -{ x_2}\,{ y_2} , { x_2}\,
 { y_3} , -{ x_1}\,{ y_1} , { x_1}\,{ y_2} , {{3\,
 { x_2}\,{ y_1}}\over{\sqrt{10}}}-{{{ x_1}\,{ y_3}}\over{
 \sqrt{10}}} \right) $$
 
$\bf2'\otimes3'=2'\oplus4'$ 
$$ {\bf 2'} \otimes {\bf 3'} : {\bf 2'} $$ 
$$\left( {{{ x_1}\,{ y_2}}\over{\sqrt{3}}}-{{2\,{ x_2}\,
 { y_1}}\over{\sqrt{6}}} , {{2\,{ x_1}\,{ y_3}}\over{\sqrt{6
 }}}-{{{ x_2}\,{ y_2}}\over{\sqrt{3}}} \right) $$
$$ {\bf 2'} \otimes {\bf 3'} : {\bf 4'} $$ 
$$\left( {{{ x_1}\,{ y_3}}\over{\sqrt{3}}}+{{2\,{ x_2}\,
 { y_2}}\over{\sqrt{6}}} , { x_1}\,{ y_1} , -{ x_2}\,
 { y_3} , {{2\,{ x_1}\,{ y_2}}\over{\sqrt{6}}}+{{{ x_2}\,
 { y_1}}\over{\sqrt{3}}} \right) $$
 
$\bf2'\otimes4=2\oplus6$ 
$$ {\bf 2'} \otimes {\bf 4} : {\bf 2} $$ 
$$\left( {{{ x_1}\,{ y_2}}\over{\sqrt{2}}}-{{{ x_2}\,
 { y_1}}\over{\sqrt{2}}} , {{{ x_1}\,{ y_4}}\over{\sqrt{2}}}
 -{{{ x_2}\,{ y_3}}\over{\sqrt{2}}} \right) $$
$$ {\bf 2'} \otimes {\bf 4} : {\bf 6} $$ 
$$\left( -{{2\,{ x_2}\,{ y_4}}\over{\sqrt{5}}}-{{{ x_1}\,
 { y_1}}\over{\sqrt{5}}} , { x_1}\,{ y_3} , {{{ x_1}\,
 { y_2}}\over{\sqrt{2}}}+{{{ x_2}\,{ y_1}}\over{\sqrt{2}}}
  , -{{{ x_1}\,{ y_4}}\over{\sqrt{2}}}-{{{ x_2}\,{ y_3}
 }\over{\sqrt{2}}} , { x_2}\,{ y_2} , {{2\,{ x_1}\,{ y_1}
 }\over{\sqrt{5}}}-{{{ x_2}\,{ y_4}}\over{\sqrt{5}}} \right) $$
 
$\bf2'\otimes4'=3'\oplus5$ 
$$ {\bf 2'} \otimes {\bf 4'} : {\bf 3'} $$ 
$$\left( {{{ x_1}\,{ y_4}}\over{2}}-{{3\,{ x_2}\,{ y_2}
 }\over{2\,\sqrt{3}}} , {{{ x_1}\,{ y_1}}\over{\sqrt{2}}}-{{
 { x_2}\,{ y_4}}\over{\sqrt{2}}} , -{{3\,{ x_1}\,{ y_3}
 }\over{2\,\sqrt{3}}}-{{{ x_2}\,{ y_1}}\over{2}} \right) $$
$$ {\bf 2'} \otimes {\bf 4'} : {\bf 5} $$ 
$$\left( {{3\,{ x_1}\,{ y_4}}\over{2\,\sqrt{3}}}+{{{ x_2}\,
 { y_2}}\over{2}} , -{ x_2}\,{ y_3} , {{{ x_2}\,{ y_4}
 }\over{\sqrt{2}}}+{{{ x_1}\,{ y_1}}\over{\sqrt{2}}} , -
 { x_1}\,{ y_2} , {{{ x_1}\,{ y_3}}\over{2}}-{{3\,
 { x_2}\,{ y_1}}\over{2\,\sqrt{3}}} \right) $$
 
$\bf2'\otimes5=4'\oplus6$ 
$$ {\bf 2'} \otimes {\bf 5} : {\bf 4'} $$ 
$$\left( {{3\,{ x_1}\,{ y_5}}\over{\sqrt{15}}}+{{2\,{ x_2}\,
 { y_3}}\over{\sqrt{10}}} , -{{2\,{ x_2}\,{ y_4}}\over{
 \sqrt{5}}}-{{{ x_1}\,{ y_1}}\over{\sqrt{5}}} , {{{ x_2}\,
 { y_5}}\over{\sqrt{5}}}+{{2\,{ x_1}\,{ y_2}}\over{\sqrt{5}
 }} , {{3\,{ x_2}\,{ y_1}}\over{\sqrt{15}}}-{{2\,{ x_1}\,
 { y_3}}\over{\sqrt{10}}} \right) $$
$$ {\bf 2'} \otimes {\bf 5} : {\bf 6} $$ 
$$\left( {{{ x_1}\,{ y_4}}\over{\sqrt{2}}}+{{{ x_2}\,
 { y_2}}\over{\sqrt{2}}} , {{2\,{ x_1}\,{ y_5}}\over{\sqrt{10
 }}}-{{3\,{ x_2}\,{ y_3}}\over{\sqrt{15}}} , {{2\,{ x_1}\,
 { y_1}}\over{\sqrt{5}}}-{{{ x_2}\,{ y_4}}\over{\sqrt{5}}}
  , {{2\,{ x_2}\,{ y_5}}\over{\sqrt{5}}}-{{{ x_1}\,{ y_2}
 }\over{\sqrt{5}}} , -{{3\,{ x_1}\,{ y_3}}\over{\sqrt{15}}}-{{2
 \,{ x_2}\,{ y_1}}\over{\sqrt{10}}} , {{{ x_1}\,{ y_4}
 }\over{\sqrt{2}}}-{{{ x_2}\,{ y_2}}\over{\sqrt{2}}} \right) $$
 
$\bf2'\otimes6=3\oplus4\oplus5$ 
$$ {\bf 2'} \otimes {\bf 6} : {\bf 3} $$ 
$$\left( -{{3\,{ x_1}\,{ y_6}}\over{2\,\sqrt{5}}}-{{{ x_2}\,
 { y_4}}\over{\sqrt{2}}}-{{{ x_1}\,{ y_1}}\over{2\,\sqrt{5}
 }} , {{{ x_2}\,{ y_5}}\over{\sqrt{2}}}+{{{ x_1}\,{ y_2}
 }\over{\sqrt{2}}} , -{{{ x_2}\,{ y_6}}\over{2\,\sqrt{5}}}-{{
 { x_1}\,{ y_3}}\over{\sqrt{2}}}+{{3\,{ x_2}\,{ y_1}
 }\over{2\,\sqrt{5}}} \right) $$
$$ {\bf 2'} \otimes {\bf 6} : {\bf 4} $$ 
$$\left( -{{4\,{ x_2}\,{ y_6}}\over{\sqrt{30}}}+{{{ x_1}\,
 { y_3}}\over{\sqrt{3}}}+{{2\,{ x_2}\,{ y_1}}\over{\sqrt{30}
 }} , {{2\,{ x_1}\,{ y_5}}\over{\sqrt{6}}}-{{{ x_2}\,
 { y_3}}\over{\sqrt{3}}} , -{{{ x_1}\,{ y_4}}\over{\sqrt{3}
 }}-{{2\,{ x_2}\,{ y_2}}\over{\sqrt{6}}} , -{{2\,{ x_1}\,
 { y_6}}\over{\sqrt{30}}}+{{{ x_2}\,{ y_4}}\over{\sqrt{3}}}-
 {{4\,{ x_1}\,{ y_1}}\over{\sqrt{30}}} \right) $$
$$ {\bf 2'} \otimes {\bf 6} : {\bf 5} $$ 
$$\left( {{{ x_1}\,{ y_5}}\over{\sqrt{3}}}+{{2\,{ x_2}\,
 { y_3}}\over{\sqrt{6}}} , {{5\,{ x_1}\,{ y_6}}\over{2\,
 \sqrt{15}}}-{{{ x_2}\,{ y_4}}\over{\sqrt{6}}}-{{5\,{ x_1}\,
 { y_1}}\over{2\,\sqrt{15}}} , {{{ x_1}\,{ y_2}}\over{\sqrt{2
 }}}-{{{ x_2}\,{ y_5}}\over{\sqrt{2}}} , {{5\,{ x_2}\,
 { y_6}}\over{2\,\sqrt{15}}}+{{{ x_1}\,{ y_3}}\over{\sqrt{6}
 }}+{{5\,{ x_2}\,{ y_1}}\over{2\,\sqrt{15}}} , {{{ x_2}\,
 { y_2}}\over{\sqrt{3}}}-{{2\,{ x_1}\,{ y_4}}\over{\sqrt{6}
 }} \right) $$
 
$\bf3\otimes3=1\oplus3\oplus5$ 
$$ {\bf 3} \otimes {\bf 3} : {\bf 1} $$ 
$$\left( { x_1}\,{ y_3}-{ x_2}\,{ y_2}+{ x_3}\,
 { y_1} \right) $$
$$ {\bf 3} \otimes {\bf 3} : {\bf 3} $$ 
$$\left( {{{ x_1}\,{ y_2}}\over{\sqrt{2}}}-{{{ x_2}\,
 { y_1}}\over{\sqrt{2}}} , {{{ x_1}\,{ y_3}}\over{\sqrt{2}}}
 -{{{ x_3}\,{ y_1}}\over{\sqrt{2}}} , {{{ x_2}\,{ y_3}
 }\over{\sqrt{2}}}-{{{ x_3}\,{ y_2}}\over{\sqrt{2}}} \right) $$
$$ {\bf 3} \otimes {\bf 3} : {\bf 5} $$ 
$$\left( { x_1}\,{ y_1} , {{{ x_1}\,{ y_2}}\over{\sqrt{2}
 }}+{{{ x_2}\,{ y_1}}\over{\sqrt{2}}} , {{{ x_1}\,{ y_3}
 }\over{\sqrt{6}}}+{{2\,{ x_2}\,{ y_2}}\over{\sqrt{6}}}+{{
 { x_3}\,{ y_1}}\over{\sqrt{6}}} , {{{ x_2}\,{ y_3}
 }\over{\sqrt{2}}}+{{{ x_3}\,{ y_2}}\over{\sqrt{2}}} , 
 { x_3}\,{ y_3} \right) $$
 
 $\bf3\otimes3'=4\oplus5$ 
$$ {\bf 3} \otimes {\bf 3'} : {\bf 4} $$ 
$$\left( {{{ x_1}\,{ y_3}}\over{\sqrt{3}}}-{{2\,{ x_3}\,
 { y_2}}\over{\sqrt{6}}} , {{{ x_3}\,{ y_3}}\over{\sqrt{3}}}
 -{{2\,{ x_2}\,{ y_1}}\over{\sqrt{6}}} , -{{2\,{ x_2}\,
 { y_3}}\over{\sqrt{6}}}-{{{ x_1}\,{ y_1}}\over{\sqrt{3}}}
  , {{2\,{ x_1}\,{ y_2}}\over{\sqrt{6}}}+{{{ x_3}\,{ y_1}
 }\over{\sqrt{3}}} \right) $$
$$ {\bf 3} \otimes {\bf 3'} : {\bf 5} $$ 
$$\left( -{{2\,{ x_3}\,{ y_3}}\over{\sqrt{6}}}-{{{ x_2}\,
 { y_1}}\over{\sqrt{3}}} , {{2\,{ x_3}\,{ y_1}}\over{\sqrt{6
 }}}-{{{ x_1}\,{ y_2}}\over{\sqrt{3}}} , { x_2}\,{ y_2}
  , -{{2\,{ x_1}\,{ y_3}}\over{\sqrt{6}}}-{{{ x_3}\,
 { y_2}}\over{\sqrt{3}}} , {{{ x_2}\,{ y_3}}\over{\sqrt{3}}}
 -{{2\,{ x_1}\,{ y_1}}\over{\sqrt{6}}} \right) $$
 
$\bf3\otimes4=3'\oplus4\oplus5$ 
$$ {\bf 3} \otimes {\bf 4} : {\bf 3'} $$ 
$$\left( -{{{ x_1}\,{ y_4}}\over{2}}+{{{ x_3}\,{ y_3}
 }\over{2}}-{{{ x_2}\,{ y_2}}\over{\sqrt{2}}} , {{{ x_1}\,
 { y_1}}\over{\sqrt{2}}}-{{{ x_3}\,{ y_4}}\over{\sqrt{2}}}
  , -{{{ x_2}\,{ y_3}}\over{\sqrt{2}}}-{{{ x_1}\,{ y_2}
 }\over{2}}-{{{ x_3}\,{ y_1}}\over{2}} \right) $$
$$ {\bf 3} \otimes {\bf 4} : {\bf 4} $$ 
$$\left( {{2\,{ x_1}\,{ y_3}}\over{\sqrt{6}}}-{{{ x_2}\,
 { y_1}}\over{\sqrt{3}}} , {{2\,{ x_1}\,{ y_4}}\over{\sqrt{6
 }}}-{{{ x_2}\,{ y_2}}\over{\sqrt{3}}} , {{{ x_2}\,{ y_3}
 }\over{\sqrt{3}}}-{{2\,{ x_3}\,{ y_1}}\over{\sqrt{6}}} , {{
 { x_2}\,{ y_4}}\over{\sqrt{3}}}-{{2\,{ x_3}\,{ y_2}
 }\over{\sqrt{6}}} \right) $$
$$ {\bf 3} \otimes {\bf 4} : {\bf 5} $$ 
$$\left( {{{ x_1}\,{ y_4}}\over{2\,\sqrt{3}}}+{{3\,{ x_3}\,
 { y_3}}\over{2\,\sqrt{3}}}+{{{ x_2}\,{ y_2}}\over{\sqrt{6}
 }} , -{{2\,{ x_2}\,{ y_4}}\over{\sqrt{6}}}-{{{ x_3}\,
 { y_2}}\over{\sqrt{3}}} , {{{ x_3}\,{ y_4}}\over{\sqrt{2}}}
 +{{{ x_1}\,{ y_1}}\over{\sqrt{2}}} , -{{{ x_1}\,{ y_3}
 }\over{\sqrt{3}}}-{{2\,{ x_2}\,{ y_1}}\over{\sqrt{6}}} , {{
 { x_2}\,{ y_3}}\over{\sqrt{6}}}-{{3\,{ x_1}\,{ y_2}
 }\over{2\,\sqrt{3}}}+{{{ x_3}\,{ y_1}}\over{2\,\sqrt{3}}}
  \right) $$
  
$\bf3\otimes4'=2\oplus4'\oplus6$ 
$$ {\bf 3} \otimes {\bf 4'} : {\bf 2} $$ 
$$\left( {{{ x_1}\,{ y_3}}\over{\sqrt{6}}}-{{{ x_2}\,
 { y_2}}\over{\sqrt{3}}}+{{{ x_3}\,{ y_1}}\over{\sqrt{2}}}
  , {{{ x_1}\,{ y_4}}\over{\sqrt{2}}}-{{{ x_2}\,{ y_3}
 }\over{\sqrt{3}}}+{{{ x_3}\,{ y_2}}\over{\sqrt{6}}} \right) $$
$$ {\bf 3} \otimes {\bf 4'} : {\bf 4'} $$ 
$$\left( {{2\,{ x_1}\,{ y_2}}\over{\sqrt{10}}}-{{3\,{ x_2}\,
 { y_1}}\over{\sqrt{15}}} , {{4\,{ x_1}\,{ y_3}}\over{\sqrt{30
 }}}-{{{ x_2}\,{ y_2}}\over{\sqrt{15}}}-{{2\,{ x_3}\,
 { y_1}}\over{\sqrt{10}}} , {{2\,{ x_1}\,{ y_4}}\over{\sqrt{10
 }}}+{{{ x_2}\,{ y_3}}\over{\sqrt{15}}}-{{4\,{ x_3}\,
 { y_2}}\over{\sqrt{30}}} , {{3\,{ x_2}\,{ y_4}}\over{\sqrt{15
 }}}-{{2\,{ x_3}\,{ y_3}}\over{\sqrt{10}}} \right) $$
$$ {\bf 3} \otimes {\bf 4'} : {\bf 6} $$ 
$$\left( { x_1}\,{ y_1} , {{3\,{ x_1}\,{ y_2}}\over{
 \sqrt{15}}}+{{2\,{ x_2}\,{ y_1}}\over{\sqrt{10}}} , {{3\,
 { x_1}\,{ y_3}}\over{\sqrt{30}}}+{{3\,{ x_2}\,{ y_2}
 }\over{\sqrt{15}}}+{{{ x_3}\,{ y_1}}\over{\sqrt{10}}} , {{
 { x_1}\,{ y_4}}\over{\sqrt{10}}}+{{3\,{ x_2}\,{ y_3}
 }\over{\sqrt{15}}}+{{3\,{ x_3}\,{ y_2}}\over{\sqrt{30}}} , {{2
 \,{ x_2}\,{ y_4}}\over{\sqrt{10}}}+{{3\,{ x_3}\,{ y_3}
 }\over{\sqrt{15}}} , { x_3}\,{ y_4} \right) $$
 
$\bf3\otimes5=(3)\oplus(5)\oplus(3'\oplus4)$ 
$$ {\bf 3} \otimes {\bf 5} : {\bf 3} $$ 
$$\left( {{{ x_1}\,{ y_3}}\over{\sqrt{10}}}-{{3\,{ x_2}\,
 { y_2}}\over{\sqrt{30}}}+{{3\,{ x_3}\,{ y_1}}\over{\sqrt{15
 }}} , {{3\,{ x_1}\,{ y_4}}\over{\sqrt{30}}}-{{2\,{ x_2}\,
 { y_3}}\over{\sqrt{10}}}+{{3\,{ x_3}\,{ y_2}}\over{\sqrt{30
 }}} , {{3\,{ x_1}\,{ y_5}}\over{\sqrt{15}}}-{{3\,{ x_2}\,
 { y_4}}\over{\sqrt{30}}}+{{{ x_3}\,{ y_3}}\over{\sqrt{10}}}
  \right) $$
$$ {\bf 3} \otimes {\bf 5} : {\bf 5} $$ 
$$\left( {{{ x_1}\,{ y_2}}\over{\sqrt{3}}}-{{2\,{ x_2}\,
 { y_1}}\over{\sqrt{6}}} , {{{ x_1}\,{ y_3}}\over{\sqrt{2}}}
 -{{{ x_2}\,{ y_2}}\over{\sqrt{6}}}-{{{ x_3}\,{ y_1}
 }\over{\sqrt{3}}} , {{{ x_1}\,{ y_4}}\over{\sqrt{2}}}-{{
 { x_3}\,{ y_2}}\over{\sqrt{2}}} , {{{ x_1}\,{ y_5}
 }\over{\sqrt{3}}}+{{{ x_2}\,{ y_4}}\over{\sqrt{6}}}-{{
 { x_3}\,{ y_3}}\over{\sqrt{2}}} , {{2\,{ x_2}\,{ y_5}
 }\over{\sqrt{6}}}-{{{ x_3}\,{ y_4}}\over{\sqrt{3}}} \right) $$
$$ {\bf 3} \otimes {\bf 5} : {\bf 3'} $$ 
$$\left( {{2\,{ x_3}\,{ y_5}}\over{\sqrt{10}}}-{{2\,{ x_1}\,
 { y_2}}\over{\sqrt{10}}}-{{{ x_2}\,{ y_1}}\over{\sqrt{5}}}
  , {{{ x_1}\,{ y_4}}\over{\sqrt{5}}}+{{3\,{ x_2}\,{ y_3}
 }\over{\sqrt{15}}}+{{{ x_3}\,{ y_2}}\over{\sqrt{5}}} , {{
 { x_2}\,{ y_5}}\over{\sqrt{5}}}+{{2\,{ x_3}\,{ y_4}
 }\over{\sqrt{10}}}+{{2\,{ x_1}\,{ y_1}}\over{\sqrt{10}}}
  \right) $$
$$
 {\bf 3} \otimes {\bf 5} : {\bf 4} $$ 
$$
\left( {{{ x_1}\,{ y_5}}\over{\sqrt{15}}}+{{4\,{ x_2}\,
 { y_4}}\over{\sqrt{30}}}+{{2\,{ x_3}\,{ y_3}}\over{\sqrt{10}}} , -{{3\,{ x_3}\,{ y_5}}\over{\sqrt{15}}}-{{2\,{ x_1}\, { y_2}}\over{\sqrt{15}}}-{{2\,{ x_2}\,{ y_1}}\over{\sqrt{30 }}} , 
 \right.
 $$
 $$
 \left.
 -{{2\,{ x_2}\,{ y_5}}\over{\sqrt{30}}}-{{2\,{ x_3}\,
 { y_4}}\over{\sqrt{15}}}+{{3\,{ x_1}\,{ y_1}}\over{\sqrt{15
 }}} , {{2\,{ x_1}\,{ y_3}}\over{\sqrt{10}}}+{{4\,{ x_2}\, \\
  { y_2}}\over{\sqrt{30}}}+{{{ x_3}\,{ y_1}}\over{\sqrt{15}}}
  \right)  $$
  
$\bf3\otimes6=(4')\oplus(6)\oplus(2'\oplus6)$ 
$$ {\bf 3} \otimes {\bf 6} : {\bf 4'} $$ 
$$\left( {{{ x_1}\,{ y_3}}\over{\sqrt{15}}}-{{2\,{ x_2}\,
 { y_2}}\over{\sqrt{15}}}+{{2\,{ x_3}\,{ y_1}}\over{\sqrt{6}
 }} , {{{ x_1}\,{ y_4}}\over{\sqrt{5}}}-{{2\,{ x_2}\,
 { y_3}}\over{\sqrt{10}}}+{{2\,{ x_3}\,{ y_2}}\over{\sqrt{10
 }}} , {{2\,{ x_1}\,{ y_5}}\over{\sqrt{10}}}-{{2\,{ x_2}\,
 { y_4}}\over{\sqrt{10}}}+{{{ x_3}\,{ y_3}}\over{\sqrt{5}}}
  , {{2\,{ x_1}\,{ y_6}}\over{\sqrt{6}}}-{{2\,{ x_2}\,
 { y_5}}\over{\sqrt{15}}}+{{{ x_3}\,{ y_4}}\over{\sqrt{15}}}
  \right) $$
$$ {\bf 3} \otimes {\bf 6} : {\bf 6} $$ 
$$\left( { x_1}\,{ y_2}-{{5\,{ x_2}\,{ y_1}}\over{\sqrt{10 }}} , 
{{4\,{ x_1}\,{ y_3}}\over{\sqrt{10}}}-{{3\,{ x_2}\, { y_2}}\over{\sqrt{10}}}-{ x_3}\,{ y_1} ,
  {{3\,{ x_1}\, { y_4}}\over{\sqrt{5}}}-{{{ x_2}\,{ y_3}}\over{\sqrt{10}}}-
 {{4\,{ x_3}\,{ y_2}}\over{\sqrt{10}}} , 
 {{4\,{ x_1}\,
 { y_5}}\over{\sqrt{10}}}+{{{ x_2}\,{ y_4}}\over{\sqrt{10}}}
 -{{3\,{ x_3}\,{ y_3}}\over{\sqrt{5}}} ,
  \right.
 $$
 $$
 \left. 
 { x_1}\,{ y_6}+{{3\,{ x_2}\,{ y_5}}\over{\sqrt{10}}}-{{4\,{ x_3}\,
 { y_4}}\over{\sqrt{10}}} , {{5\,{ x_2}\,{ y_6}}\over{\sqrt{10
 }}}-{ x_3}\,{ y_5} \right) $$
$$ {\bf 3} \otimes {\bf 6} : {\bf 2'} $$ 
$$\left( -{{{ x_1}\,{ y_6}}\over{\sqrt{30}}}-{{{ x_2}\,
 { y_5}}\over{\sqrt{3}}}-{{{ x_3}\,{ y_4}}\over{\sqrt{3}}}+
 {{3\,{ x_1}\,{ y_1}}\over{\sqrt{30}}} , {{3\,{ x_3}\,
 { y_6}}\over{\sqrt{30}}}+{{{ x_1}\,{ y_3}}\over{\sqrt{3}}}+
 {{{ x_2}\,{ y_2}}\over{\sqrt{3}}}+{{{ x_3}\,{ y_1}
 }\over{\sqrt{30}}} \right) $$

$$ {\bf 3} \otimes {\bf 6} : {\bf 6} $$ 
$$\left( -{{7\,{ x_2}\,{ y_6}}\over{5\,\sqrt{5}}}-{{7\,
 { x_3}\,{ y_5}}\over{5\,\sqrt{2}}}+{{{ x_1}\,{ y_2}
 }\over{5\,\sqrt{2}}}+{{{ x_2}\,{ y_1}}\over{5\,\sqrt{5}}} , {{7
 \,{ x_3}\,{ y_6}}\over{5\,\sqrt{2}}}-{{{ x_1}\,{ y_3}
 }\over{\sqrt{5}}}-{{{ x_2}\,{ y_2}}\over{\sqrt{5}}}-{{
 { x_3}\,{ y_1}}\over{5\,\sqrt{2}}} , {{2\,{ x_1}\,{ y_4}
 }\over{\sqrt{10}}}+{{2\,{ x_2}\,{ y_3}}\over{\sqrt{5}}}+{{
 { x_3}\,{ y_2}}\over{\sqrt{5}}} ,
  \right.
 $$
 $$
 \left.
  -{{{ x_1}\,{ y_5}
 }\over{\sqrt{5}}}-{{2\,{ x_2}\,{ y_4}}\over{\sqrt{5}}}-{{2\,
 { x_3}\,{ y_3}}\over{\sqrt{10}}} , 
 {{{ x_1}\,{ y_6}
 }\over{5\,\sqrt{2}}}+{{{ x_2}\,{ y_5}}\over{\sqrt{5}}}+{{
 { x_3}\,{ y_4}}\over{\sqrt{5}}}+{{7\,{ x_1}\,{ y_1}
 }\over{5\,\sqrt{2}}} , -{{{ x_2}\,{ y_6}}\over{5\,\sqrt{5}}}-
 {{{ x_3}\,{ y_5}}\over{5\,\sqrt{2}}}-{{7\,{ x_1}\,{ y_2}
 }\over{5\,\sqrt{2}}}-{{7\,{ x_2}\,{ y_1}}\over{5\,\sqrt{5}}}
  \right) $$
  
$\bf3'\otimes3'=(1)\oplus(3')\oplus(5)$ 
$$ {\bf 3'} \otimes {\bf 3'} : {\bf 1} $$ 
$$\left( { x_1}\,{ y_3}-{ x_2}\,{ y_2}+{ x_3}\,
 { y_1} \right) $$
$$ {\bf 3'} \otimes {\bf 3'} : {\bf 3'} $$ 
$$\left( {{{ x_1}\,{ y_2}}\over{\sqrt{2}}}-{{{ x_2}\,
 { y_1}}\over{\sqrt{2}}} , {{{ x_1}\,{ y_3}}\over{\sqrt{2}}}
 -{{{ x_3}\,{ y_1}}\over{\sqrt{2}}} , {{{ x_2}\,{ y_3}
 }\over{\sqrt{2}}}-{{{ x_3}\,{ y_2}}\over{\sqrt{2}}} \right) $$
$$ {\bf 3'} \otimes {\bf 3'} : {\bf 5} $$ 
$$\left( {{{ x_1}\,{ y_2}}\over{\sqrt{2}}}+{{{ x_2}\,
 { y_1}}\over{\sqrt{2}}} , { x_3}\,{ y_3} , {{{ x_1}\,
 { y_3}}\over{\sqrt{6}}}+{{2\,{ x_2}\,{ y_2}}\over{\sqrt{6}
 }}+{{{ x_3}\,{ y_1}}\over{\sqrt{6}}} , -{ x_1}\,{ y_1}
  , -{{{ x_2}\,{ y_3}}\over{\sqrt{2}}}-{{{ x_3}\,{ y_2}
 }\over{\sqrt{2}}} \right) $$
 
$\bf3'\otimes4=3\oplus4\oplus5$ 
$$ {\bf 3'} \otimes {\bf 4} : {\bf 3} $$ 
$$\left( {{{ x_2}\,{ y_4}}\over{\sqrt{2}}}+{{{ x_3}\,
 { y_3}}\over{2}}-{{{ x_1}\,{ y_1}}\over{2}} , {{{ x_1}\,
 { y_3}}\over{\sqrt{2}}}+{{{ x_3}\,{ y_2}}\over{\sqrt{2}}}
  , -{{{ x_3}\,{ y_4}}\over{2}}-{{{ x_1}\,{ y_2}}\over{2
 }}-{{{ x_2}\,{ y_1}}\over{\sqrt{2}}} \right) $$
$$ {\bf 3'} \otimes {\bf 4} : {\bf 4} $$ 
$$\left( {{2\,{ x_1}\,{ y_2}}\over{\sqrt{6}}}-{{{ x_2}\,
 { y_1}}\over{\sqrt{3}}} , {{{ x_2}\,{ y_2}}\over{\sqrt{3}}}
 -{{2\,{ x_3}\,{ y_1}}\over{\sqrt{6}}} , {{2\,{ x_1}\,
 { y_4}}\over{\sqrt{6}}}-{{{ x_2}\,{ y_3}}\over{\sqrt{3}}}
  , {{{ x_2}\,{ y_4}}\over{\sqrt{3}}}-{{2\,{ x_3}\,{ y_3}
 }\over{\sqrt{6}}} \right) $$
$$ {\bf 3'} \otimes {\bf 4} : {\bf 5} $$ 
$$\left( {{2\,{ x_2}\,{ y_2}}\over{\sqrt{6}}}+{{{ x_3}\,
 { y_1}}\over{\sqrt{3}}} , {{{ x_2}\,{ y_4}}\over{\sqrt{6}}}
 +{{{ x_3}\,{ y_3}}\over{2\,\sqrt{3}}}+{{3\,{ x_1}\,
 { y_1}}\over{2\,\sqrt{3}}} , {{{ x_1}\,{ y_3}}\over{\sqrt{2
 }}}-{{{ x_3}\,{ y_2}}\over{\sqrt{2}}} , -{{3\,{ x_3}\,
 { y_4}}\over{2\,\sqrt{3}}}+{{{ x_1}\,{ y_2}}\over{2\,\sqrt{3
 }}}+{{{ x_2}\,{ y_1}}\over{\sqrt{6}}} , {{{ x_1}\,{ y_4}
 }\over{\sqrt{3}}}+{{2\,{ x_2}\,{ y_3}}\over{\sqrt{6}}}
  \right) $$
  
$\bf3'\otimes4'=2'\oplus4'\oplus6$ 
$$ {\bf 3'} \otimes {\bf 4'} : {\bf 2'} $$ 
$$\left( -{{{ x_2}\,{ y_4}}\over{\sqrt{3}}}+{{{ x_3}\,
 { y_2}}\over{\sqrt{2}}}+{{{ x_1}\,{ y_1}}\over{\sqrt{6}}}
  , {{{ x_3}\,{ y_4}}\over{\sqrt{6}}}-{{{ x_1}\,{ y_3}
 }\over{\sqrt{2}}}-{{{ x_2}\,{ y_1}}\over{\sqrt{3}}} \right) $$
$$ {\bf 3'} \otimes {\bf 4'} : {\bf 4'} $$ 
$$\left( {{4\,{ x_3}\,{ y_4}}\over{\sqrt{30}}}+{{2\,{ x_1}\,
 { y_3}}\over{\sqrt{10}}}-{{{ x_2}\,{ y_1}}\over{\sqrt{15}}}
  , {{3\,{ x_2}\,{ y_2}}\over{\sqrt{15}}}-{{2\,{ x_1}\,
 { y_4}}\over{\sqrt{10}}} , -{{3\,{ x_2}\,{ y_3}}\over{
 \sqrt{15}}}-{{2\,{ x_3}\,{ y_1}}\over{\sqrt{10}}} , {{
 { x_2}\,{ y_4}}\over{\sqrt{15}}}+{{2\,{ x_3}\,{ y_2}
 }\over{\sqrt{10}}}-{{4\,{ x_1}\,{ y_1}}\over{\sqrt{30}}}
  \right) $$
$$ {\bf 3'} \otimes {\bf 4'} : {\bf 6} $$ 
$$\left( {{{ x_3}\,{ y_3}}\over{\sqrt{2}}}+{{{ x_1}\,
 { y_2}}\over{\sqrt{2}}} , {{3\,{ x_3}\,{ y_4}}\over{\sqrt{30
 }}}-{{{ x_1}\,{ y_3}}\over{\sqrt{10}}}+{{3\,{ x_2}\,
 { y_1}}\over{\sqrt{15}}} , -{{3\,{ x_1}\,{ y_4}}\over{
 \sqrt{15}}}-{{2\,{ x_2}\,{ y_2}}\over{\sqrt{10}}} , {{3\,
 { x_3}\,{ y_1}}\over{\sqrt{15}}}-{{2\,{ x_2}\,{ y_3}
 }\over{\sqrt{10}}} ,
  \right.
 $$
 $$
 \left.
  {{3\,{ x_2}\,{ y_4}}\over{\sqrt{15}}}+{{
 { x_3}\,{ y_2}}\over{\sqrt{10}}}+{{3\,{ x_1}\,{ y_1}
 }\over{\sqrt{30}}} , {{{ x_1}\,{ y_2}}\over{\sqrt{2}}}-{{
 { x_3}\,{ y_3}}\over{\sqrt{2}}} \right) $$
 
$\bf3'\otimes5=(3')\oplus(5)\oplus(3\oplus4)$ 
$$ {\bf 3'} \otimes {\bf 5} : {\bf 3'} $$ 
$$\left( -{{3\,{ x_3}\,{ y_4}}\over{\sqrt{15}}}+{{{ x_1}\,
 { y_3}}\over{\sqrt{10}}}-{{3\,{ x_2}\,{ y_1}}\over{\sqrt{30
 }}} , -{{3\,{ x_1}\,{ y_5}}\over{\sqrt{30}}}-{{2\,{ x_2}\,
 { y_3}}\over{\sqrt{10}}}+{{3\,{ x_3}\,{ y_1}}\over{\sqrt{30
 }}} , {{3\,{ x_2}\,{ y_5}}\over{\sqrt{30}}}+{{{ x_3}\,
 { y_3}}\over{\sqrt{10}}}+{{3\,{ x_1}\,{ y_2}}\over{\sqrt{15
 }}} \right) $$
$$ {\bf 3'} \otimes {\bf 5} : {\bf 5} $$ 
$$\left( {{{ x_3}\,{ y_4}}\over{\sqrt{3}}}+{{{ x_1}\,
 { y_3}}\over{\sqrt{2}}}-{{{ x_2}\,{ y_1}}\over{\sqrt{6}}}
  , {{{ x_3}\,{ y_5}}\over{\sqrt{3}}}+{{2\,{ x_2}\,{ y_2}
 }\over{\sqrt{6}}} , -{{{ x_1}\,{ y_5}}\over{\sqrt{2}}}-{{
 { x_3}\,{ y_1}}\over{\sqrt{2}}} , -{{2\,{ x_2}\,{ y_4}
 }\over{\sqrt{6}}}-{{{ x_1}\,{ y_1}}\over{\sqrt{3}}} , {{
 { x_2}\,{ y_5}}\over{\sqrt{6}}}+{{{ x_3}\,{ y_3}}\over{
 \sqrt{2}}}-{{{ x_1}\,{ y_2}}\over{\sqrt{3}}} \right) $$
$$ {\bf 3'} \otimes {\bf 5} : {\bf 3} $$ 
$$\left( {{2\,{ x_3}\,{ y_5}}\over{\sqrt{10}}}+{{2\,{ x_1}\,
 { y_4}}\over{\sqrt{10}}}-{{{ x_2}\,{ y_2}}\over{\sqrt{5}}}
  , -{{{ x_1}\,{ y_5}}\over{\sqrt{5}}}+{{3\,{ x_2}\,
 { y_3}}\over{\sqrt{15}}}+{{{ x_3}\,{ y_1}}\over{\sqrt{5}}}
  , -{{{ x_2}\,{ y_4}}\over{\sqrt{5}}}-{{2\,{ x_3}\,
 { y_2}}\over{\sqrt{10}}}+{{2\,{ x_1}\,{ y_1}}\over{\sqrt{10
 }}} \right) $$
$$ {\bf 3'} \otimes {\bf 5} : {\bf 4} $$ 
$$\left( -{{2\,{ x_2}\,{ y_4}}\over{\sqrt{30}}}+{{3\,{ x_3}
 \,{ y_2}}\over{\sqrt{15}}}+{{2\,{ x_1}\,{ y_1}}\over{\sqrt{15
 }}} , {{{ x_3}\,{ y_4}}\over{\sqrt{15}}}-{{2\,{ x_1}\,
 { y_3}}\over{\sqrt{10}}}-{{4\,{ x_2}\,{ y_1}}\over{\sqrt{30
 }}} , 
  \right.
 $$
 $$
 \left.
 -{{4\,{ x_2}\,{ y_5}}\over{\sqrt{30}}}+{{2\,{ x_3}\,
 { y_3}}\over{\sqrt{10}}}+{{{ x_1}\,{ y_2}}\over{\sqrt{15}}}
  , {{2\,{ x_3}\,{ y_5}}\over{\sqrt{15}}}-{{3\,{ x_1}\,
 { y_4}}\over{\sqrt{15}}}-{{2\,{ x_2}\,{ y_2}}\over{\sqrt{30
 }}} \right) $$

$\bf3'\otimes6=(4')\oplus(6)\oplus(2\oplus6)$ 
$$ {\bf 3'} \otimes {\bf 6} : {\bf 4'} $$ 
$$\left( {{{ x_3}\,{ y_5}}\over{\sqrt{5}}}+{{2\,{ x_1}\,
 { y_4}}\over{\sqrt{10}}}-{{2\,{ x_2}\,{ y_2}}\over{\sqrt{10
 }}} , {{{ x_3}\,{ y_6}}\over{\sqrt{3}}}+{{{ x_1}\,{ y_5}
 }\over{\sqrt{15}}}+{{2\,{ x_2}\,{ y_3}}\over{\sqrt{15}}}+{{
 { x_3}\,{ y_1}}\over{\sqrt{3}}} ,
  \right.
 $$
 $$
 \left.
  -{{{ x_1}\,{ y_6}
 }\over{\sqrt{3}}}+{{2\,{ x_2}\,{ y_4}}\over{\sqrt{15}}}-{{
 { x_3}\,{ y_2}}\over{\sqrt{15}}}+{{{ x_1}\,{ y_1}}\over{
 \sqrt{3}}} , -{{2\,{ x_2}\,{ y_5}}\over{\sqrt{10}}}-{{2\,
 { x_3}\,{ y_3}}\over{\sqrt{10}}}+{{{ x_1}\,{ y_2}}\over{
 \sqrt{5}}} \right) $$
$$ {\bf 3'} \otimes {\bf 6} : {\bf 6} $$ 
$$\left( {{5\,{ x_2}\,{ y_6}}\over{\sqrt{10}}}-{{{ x_3}\,
 { y_4}}\over{\sqrt{2}}}+{{{ x_1}\,{ y_3}}\over{\sqrt{2}}}
  , {{3\,{ x_3}\,{ y_5}}\over{\sqrt{5}}}-{{4\,{ x_1}\,
 { y_4}}\over{\sqrt{10}}}-{{{ x_2}\,{ y_2}}\over{\sqrt{10}}}
  , -{{{ x_3}\,{ y_6}}\over{\sqrt{2}}}+{{4\,{ x_1}\,
 { y_5}}\over{\sqrt{10}}}+{{3\,{ x_2}\,{ y_3}}\over{\sqrt{10
 }}}-{{{ x_3}\,{ y_1}}\over{\sqrt{2}}} ,
  \right.
 $$
 $$
 \left.
  -{{{ x_1}\,
 { y_6}}\over{\sqrt{2}}}-{{3\,{ x_2}\,{ y_4}}\over{\sqrt{10}
 }}+{{4\,{ x_3}\,{ y_2}}\over{\sqrt{10}}}+{{{ x_1}\,
 { y_1}}\over{\sqrt{2}}} , {{{ x_2}\,{ y_5}}\over{\sqrt{10}
 }}-{{4\,{ x_3}\,{ y_3}}\over{\sqrt{10}}}-{{3\,{ x_1}\,
 { y_2}}\over{\sqrt{5}}} , {{{ x_3}\,{ y_4}}\over{\sqrt{2}}}
 +{{{ x_1}\,{ y_3}}\over{\sqrt{2}}}+{{5\,{ x_2}\,{ y_1}
 }\over{\sqrt{10}}} \right) $$
$$ {\bf 3'} \otimes {\bf 6} : {\bf 2} $$ 
$$\left( {{2\,{ x_3}\,{ y_6}}\over{\sqrt{15}}}+{{{ x_1}\,
 { y_5}}\over{\sqrt{3}}}-{{{ x_2}\,{ y_3}}\over{\sqrt{3}}}-
 {{{ x_3}\,{ y_1}}\over{\sqrt{15}}} , -{{{ x_1}\,{ y_6}
 }\over{\sqrt{15}}}+{{{ x_2}\,{ y_4}}\over{\sqrt{3}}}+{{
 { x_3}\,{ y_2}}\over{\sqrt{3}}}-{{2\,{ x_1}\,{ y_1}
 }\over{\sqrt{15}}} \right) $$

$$ {\bf 3'} \otimes {\bf 6} : {\bf 6} $$ 
$$\left( {{{ x_2}\,{ y_6}}\over{5\,\sqrt{5}}}-{{3\,{ x_3}\,
 { y_4}}\over{5}}-{{4\,{ x_1}\,{ y_3}}\over{5}}+{{7\,
 { x_2}\,{ y_1}}\over{5\,\sqrt{5}}} , -{{2\,{ x_3}\,
 { y_5}}\over{\sqrt{10}}}-{{{ x_1}\,{ y_4}}\over{\sqrt{5}}}-
 {{2\,{ x_2}\,{ y_2}}\over{\sqrt{5}}} , -{{3\,{ x_3}\,
 { y_6}}\over{5}}+{{{ x_1}\,{ y_5}}\over{\sqrt{5}}}-{{
 { x_2}\,{ y_3}}\over{\sqrt{5}}}+{{4\,{ x_3}\,{ y_1}
 }\over{5}} , 
  \right.
 $$
 $$
 \left.
 {{4\,{ x_1}\,{ y_6}}\over{5}}+{{{ x_2}\,
 { y_4}}\over{\sqrt{5}}}+{{{ x_3}\,{ y_2}}\over{\sqrt{5}}}+
 {{3\,{ x_1}\,{ y_1}}\over{5}} , {{2\,{ x_2}\,{ y_5}
 }\over{\sqrt{5}}}-{{{ x_3}\,{ y_3}}\over{\sqrt{5}}}+{{2\,
 { x_1}\,{ y_2}}\over{\sqrt{10}}} , -{{7\,{ x_2}\,{ y_6}
 }\over{5\,\sqrt{5}}}-{{4\,{ x_3}\,{ y_4}}\over{5}}+{{3\,
 { x_1}\,{ y_3}}\over{5}}+{{{ x_2}\,{ y_1}}\over{5\,
 \sqrt{5}}} \right) $$
 
$\bf4\otimes4=1\oplus3\oplus3'\oplus4\oplus5$ 
$$ {\bf 4} \otimes {\bf 4} : {\bf 1} $$ 
$$\left( { x_1}\,{ y_4}-{ x_2}\,{ y_3}-{ x_3}\,
 { y_2}+{ x_4}\,{ y_1} \right) $$
$$ {\bf 4} \otimes {\bf 4} : {\bf 3} $$ 
$$\left( {{{ x_1}\,{ y_2}}\over{\sqrt{2}}}-{{{ x_2}\,
 { y_1}}\over{\sqrt{2}}} , {{{ x_1}\,{ y_4}}\over{2}}-{{
 { x_2}\,{ y_3}}\over{2}}+{{{ x_3}\,{ y_2}}\over{2}}-{{
 { x_4}\,{ y_1}}\over{2}} , {{{ x_3}\,{ y_4}}\over{\sqrt{2
 }}}-{{{ x_4}\,{ y_3}}\over{\sqrt{2}}} \right) $$
$$ {\bf 4} \otimes {\bf 4} : {\bf 3'} $$ 
$$\left( {{{ x_1}\,{ y_3}}\over{\sqrt{2}}}-{{{ x_3}\,
 { y_1}}\over{\sqrt{2}}} , {{{ x_1}\,{ y_4}}\over{2}}+{{
 { x_2}\,{ y_3}}\over{2}}-{{{ x_3}\,{ y_2}}\over{2}}-{{
 { x_4}\,{ y_1}}\over{2}} , {{{ x_2}\,{ y_4}}\over{\sqrt{2
 }}}-{{{ x_4}\,{ y_2}}\over{\sqrt{2}}} \right) $$
$$ {\bf 4} \otimes {\bf 4} : {\bf 4} $$ 
$$\left( -{{{ x_3}\,{ y_4}}\over{\sqrt{3}}}-{{{ x_4}\,
 { y_3}}\over{\sqrt{3}}}+{{{ x_2}\,{ y_2}}\over{\sqrt{3}}}
  , {{{ x_4}\,{ y_4}}\over{\sqrt{3}}}-{{{ x_1}\,{ y_3}
 }\over{\sqrt{3}}}-{{{ x_3}\,{ y_1}}\over{\sqrt{3}}} , -{{
 { x_2}\,{ y_4}}\over{\sqrt{3}}}-{{{ x_4}\,{ y_2}}\over{
 \sqrt{3}}}-{{{ x_1}\,{ y_1}}\over{\sqrt{3}}} , {{{ x_3}\,
 { y_3}}\over{\sqrt{3}}}+{{{ x_1}\,{ y_2}}\over{\sqrt{3}}}+
 {{{ x_2}\,{ y_1}}\over{\sqrt{3}}} \right) $$
$$ {\bf 4} \otimes {\bf 4} : {\bf 5} $$ 
$$\left( -{{2\,{ x_4}\,{ y_4}}\over{\sqrt{6}}}-{{{ x_1}\,
 { y_3}}\over{\sqrt{6}}}-{{{ x_3}\,{ y_1}}\over{\sqrt{6}}}
  , {{2\,{ x_3}\,{ y_3}}\over{\sqrt{6}}}-{{{ x_1}\,{ y_2}
 }\over{\sqrt{6}}}-{{{ x_2}\,{ y_1}}\over{\sqrt{6}}} , {{
 { x_1}\,{ y_4}}\over{2}}+{{{ x_2}\,{ y_3}}\over{2}}+{{
 { x_3}\,{ y_2}}\over{2}}+{{{ x_4}\,{ y_1}}\over{2}} ,
  \right.
 $$
 $$
 \left.
  - {{{ x_3}\,{ y_4}}\over{\sqrt{6}}}-{{{ x_4}\,{ y_3}
 }\over{\sqrt{6}}}-{{2\,{ x_2}\,{ y_2}}\over{\sqrt{6}}} , {{
 { x_2}\,{ y_4}}\over{\sqrt{6}}}+{{{ x_4}\,{ y_2}}\over{
 \sqrt{6}}}-{{2\,{ x_1}\,{ y_1}}\over{\sqrt{6}}} \right) $$
 
$\bf4\otimes4'=4\oplus6\oplus6$ 
$$ {\bf 4} \otimes {\bf 4'} : {\bf 4'} $$ 
$$\left( {{3\,{ x_3}\,{ y_4}}\over{\sqrt{15}}}+{{{ x_2}\,
 { y_3}}\over{\sqrt{5}}}+{{{ x_4}\,{ y_2}}\over{\sqrt{5}}}
  , -{{{ x_2}\,{ y_4}}\over{\sqrt{5}}}-{{3\,{ x_4}\,
 { y_3}}\over{\sqrt{15}}}-{{{ x_1}\,{ y_1}}\over{\sqrt{5}}}
  , {{{ x_4}\,{ y_4}}\over{\sqrt{5}}}+{{3\,{ x_1}\,{ y_2}
 }\over{\sqrt{15}}}+{{{ x_3}\,{ y_1}}\over{\sqrt{5}}} , -{{
 { x_1}\,{ y_3}}\over{\sqrt{5}}}-{{{ x_3}\,{ y_2}}\over{
 \sqrt{5}}}+{{3\,{ x_2}\,{ y_1}}\over{\sqrt{15}}} \right) $$

$$ {\bf 4} \otimes {\bf 4'} : {\bf 6} $$ 
$$\left( {{3\,\sqrt{3}\,{ x_1}\,{ y_4}}\over{2\,\sqrt{10}}}-{{3
 \,{ x_3}\,{ y_3}}\over{2\,\sqrt{10}}}+{{{ x_2}\,{ y_2}
 }\over{2\,\sqrt{10}}}-{{\sqrt{3}\,{ x_4}\,{ y_1}}\over{2\,
 \sqrt{10}}} , {{{ x_4}\,{ y_2}}\over{\sqrt{2}}}-{{{ x_2}\,
 { y_3}}\over{\sqrt{2}}} , {{3\,{ x_2}\,{ y_4}}\over{2\,
 \sqrt{3}}}-{{{ x_4}\,{ y_3}}\over{2}} , {{3\,{ x_3}\,
 { y_1}}\over{2\,\sqrt{3}}}-{{{ x_1}\,{ y_2}}\over{2}} ,
  \right.
 $$
 $$
 \left.
  {{
 { x_1}\,{ y_3}}\over{\sqrt{2}}}-{{{ x_3}\,{ y_2}}\over{
 \sqrt{2}}} , -{{\sqrt{3}\,{ x_1}\,{ y_4}}\over{2\,\sqrt{10}}}+
 {{{ x_3}\,{ y_3}}\over{2\,\sqrt{10}}}+{{3\,{ x_2}\,
 { y_2}}\over{2\,\sqrt{10}}}-{{3\,\sqrt{3}\,{ x_4}\,{ y_1}
 }\over{2\,\sqrt{10}}} \right) $$
$$ {\bf 4} \otimes {\bf 4'} : {\bf 6} $$ 
$$\left( {{{ x_1}\,{ y_4}}\over{2\,\sqrt{2}}}+{{3\,{ x_3}\,
 { y_3}}\over{2\,\sqrt{6}}}+{{3\,{ x_2}\,{ y_2}}\over{2\,
 \sqrt{6}}}+{{{ x_4}\,{ y_1}}\over{2\,\sqrt{2}}} , {{2\,
 { x_3}\,{ y_4}}\over{\sqrt{10}}}-{{3\,{ x_2}\,{ y_3}
 }\over{\sqrt{30}}}-{{3\,{ x_4}\,{ y_2}}\over{\sqrt{30}}} , -{{
 { x_2}\,{ y_4}}\over{2\,\sqrt{5}}}-{{3\,{ x_4}\,{ y_3}
 }\over{2\,\sqrt{15}}}+{{2\,{ x_1}\,{ y_1}}\over{\sqrt{5}}} , 
  \right.
 $$
 $$
 \left.
 {{2\,{ x_4}\,{ y_4}}\over{\sqrt{5}}}-{{3\,{ x_1}\,{ y_2}
 }\over{2\,\sqrt{15}}}-{{{ x_3}\,{ y_1}}\over{2\,\sqrt{5}}} , -
 {{3\,{ x_1}\,{ y_3}}\over{\sqrt{30}}}-{{3\,{ x_3}\,
 { y_2}}\over{\sqrt{30}}}-{{2\,{ x_2}\,{ y_1}}\over{\sqrt{10
 }}} , {{{ x_1}\,{ y_4}}\over{2\,\sqrt{2}}}+{{3\,{ x_3}\,
 { y_3}}\over{2\,\sqrt{6}}}-{{3\,{ x_2}\,{ y_2}}\over{2\,
 \sqrt{6}}}-{{{ x_4}\,{ y_1}}\over{2\,\sqrt{2}}} \right) $$

$\bf4\otimes5=(3'\oplus5)\oplus(3\oplus4\oplus5)$ 
$$ {\bf 4} \otimes {\bf 5} : {\bf 3'} $$ 
$$\left( {{{ x_1}\,{ y_5}}\over{\sqrt{5}}}-{{{ x_3}\,
 { y_4}}\over{2\,\sqrt{5}}}-{{3\,{ x_2}\,{ y_3}}\over{\sqrt{30
 }}}+{{3\,{ x_4}\,{ y_2}}\over{2\,\sqrt{5}}} , -{{2\,{ x_2}
 \,{ y_5}}\over{\sqrt{10}}}+{{{ x_4}\,{ y_4}}\over{\sqrt{10}
 }}+{{{ x_1}\,{ y_2}}\over{\sqrt{10}}}-{{2\,{ x_3}\,
 { y_1}}\over{\sqrt{10}}} , 
  \right.
 $$
 $$
 \left.
 -{{3\,{ x_1}\,{ y_4}}\over{2\,
 \sqrt{5}}}+{{3\,{ x_3}\,{ y_3}}\over{\sqrt{30}}}-{{{ x_2}\,
 { y_2}}\over{2\,\sqrt{5}}}+{{{ x_4}\,{ y_1}}\over{\sqrt{5}
 }} \right) $$
$$ {\bf 4} \otimes {\bf 5} : {\bf 5} $$ 
$$\left( {{3\,{ x_1}\,{ y_5}}\over{\sqrt{15}}}-{{3\,{ x_3}\,
 { y_4}}\over{2\,\sqrt{15}}}+{{{ x_2}\,{ y_3}}\over{\sqrt{10
 }}}-{{3\,{ x_4}\,{ y_2}}\over{2\,\sqrt{15}}} , {{2\,{ x_4}
 \,{ y_3}}\over{\sqrt{10}}}-{{3\,{ x_2}\,{ y_4}}\over{\sqrt{15
 }}} , {{2\,{ x_2}\,{ y_5}}\over{\sqrt{10}}}-{{{ x_4}\,
 { y_4}}\over{\sqrt{10}}}+{{{ x_1}\,{ y_2}}\over{\sqrt{10}}}
 -{{2\,{ x_3}\,{ y_1}}\over{\sqrt{10}}} , 
  \right.
 $$
 $$
 \left.
 {{3\,{ x_3}\,
 { y_2}}\over{\sqrt{15}}}-{{2\,{ x_1}\,{ y_3}}\over{\sqrt{10
 }}} , {{3\,{ x_1}\,{ y_4}}\over{2\,\sqrt{15}}}-{{{ x_3}\,
 { y_3}}\over{\sqrt{10}}}-{{3\,{ x_2}\,{ y_2}}\over{2\,
 \sqrt{15}}}+{{3\,{ x_4}\,{ y_1}}\over{\sqrt{15}}} \right) $$
$$ {\bf 4} \otimes {\bf 5} : {\bf 3} $$ 
$$\left( -{{3\,{ x_3}\,{ y_5}}\over{2\,\sqrt{5}}}-{{{ x_2}\,
 { y_4}}\over{\sqrt{5}}}-{{3\,{ x_4}\,{ y_3}}\over{\sqrt{30}
 }}-{{{ x_1}\,{ y_1}}\over{2\,\sqrt{5}}} , {{{ x_2}\,
 { y_5}}\over{\sqrt{10}}}+{{2\,{ x_4}\,{ y_4}}\over{\sqrt{10
 }}}+{{2\,{ x_1}\,{ y_2}}\over{\sqrt{10}}}+{{{ x_3}\,
 { y_1}}\over{\sqrt{10}}} , 
  \right.
 $$
 $$
 \left.
 -{{{ x_4}\,{ y_5}}\over{2\,
 \sqrt{5}}}-{{3\,{ x_1}\,{ y_3}}\over{\sqrt{30}}}-{{{ x_3}\,
 { y_2}}\over{\sqrt{5}}}+{{3\,{ x_2}\,{ y_1}}\over{2\,\sqrt{5
 }}} \right) $$
$$ {\bf 4} \otimes {\bf 5} : {\bf 4} $$ 
$$\left( -{{4\,{ x_4}\,{ y_5}}\over{\sqrt{30}}}+{{{ x_1}\,
 { y_3}}\over{\sqrt{5}}}+{{2\,{ x_3}\,{ y_2}}\over{\sqrt{30}
 }}+{{2\,{ x_2}\,{ y_1}}\over{\sqrt{30}}} , {{2\,{ x_1}\,
 { y_5}}\over{\sqrt{30}}}+{{4\,{ x_3}\,{ y_4}}\over{\sqrt{30
 }}}-{{{ x_2}\,{ y_3}}\over{\sqrt{5}}}-{{2\,{ x_4}\,
 { y_2}}\over{\sqrt{30}}} , 
  \right.
 $$
 $$
 \left.
 -{{2\,{ x_1}\,{ y_4}}\over{
 \sqrt{30}}}-{{{ x_3}\,{ y_3}}\over{\sqrt{5}}}-{{4\,{ x_2}\,
 { y_2}}\over{\sqrt{30}}}-{{2\,{ x_4}\,{ y_1}}\over{\sqrt{30
 }}} , -{{2\,{ x_3}\,{ y_5}}\over{\sqrt{30}}}+{{2\,{ x_2}\,
 { y_4}}\over{\sqrt{30}}}+{{{ x_4}\,{ y_3}}\over{\sqrt{5}}}-
 {{4\,{ x_1}\,{ y_1}}\over{\sqrt{30}}} \right) $$
$$ {\bf 4} \otimes {\bf 5} : {\bf 5} $$ 
$$\left( {{{ x_1}\,{ y_5}}\over{\sqrt{15}}}+{{2\,{ x_3}\,
 { y_4}}\over{\sqrt{15}}}+{{2\,{ x_2}\,{ y_3}}\over{\sqrt{10
 }}}+{{2\,{ x_4}\,{ y_2}}\over{\sqrt{15}}} , {{5\,{ x_3}\,
 { y_5}}\over{2\,\sqrt{15}}}-{{{ x_2}\,{ y_4}}\over{\sqrt{15
 }}}-{{{ x_4}\,{ y_3}}\over{\sqrt{10}}}-{{5\,{ x_1}\,
 { y_1}}\over{2\,\sqrt{15}}} , 
  \right.
 $$
 $$
 \left.
 -{{{ x_2}\,{ y_5}}\over{
 \sqrt{10}}}-{{2\,{ x_4}\,{ y_4}}\over{\sqrt{10}}}+{{2\,
 { x_1}\,{ y_2}}\over{\sqrt{10}}}+{{{ x_3}\,{ y_1}}\over{
 \sqrt{10}}} , 
  \right.
 $$
 $$
 \left.
 {{5\,{ x_4}\,{ y_5}}\over{2\,\sqrt{15}}}+{{
 { x_1}\,{ y_3}}\over{\sqrt{10}}}+{{{ x_3}\,{ y_2}}\over{
 \sqrt{15}}}+{{5\,{ x_2}\,{ y_1}}\over{2\,\sqrt{15}}} , -{{2\,
 { x_1}\,{ y_4}}\over{\sqrt{15}}}-{{2\,{ x_3}\,{ y_3}
 }\over{\sqrt{10}}}+{{2\,{ x_2}\,{ y_2}}\over{\sqrt{15}}}+{{
 { x_4}\,{ y_1}}\over{\sqrt{15}}} \right) $$

$\bf4\otimes6=(2'\oplus4')\oplus(2\oplus6)\oplus(4'\oplus6)$ 
$$ {\bf 4} \otimes {\bf 6} : {\bf 2'} $$ 
$$\left( -{{2\,{ x_4}\,{ y_6}}\over{\sqrt{15}}}+{{{ x_1}\,
 { y_4}}\over{\sqrt{6}}}+{{{ x_3}\,{ y_3}}\over{\sqrt{6}}}+
 {{{ x_2}\,{ y_2}}\over{\sqrt{3}}}+{{{ x_4}\,{ y_1}
 }\over{\sqrt{15}}} , {{{ x_1}\,{ y_6}}\over{\sqrt{15}}}+{{
 { x_3}\,{ y_5}}\over{\sqrt{3}}}-{{{ x_2}\,{ y_4}}\over{
 \sqrt{6}}}-{{{ x_4}\,{ y_3}}\over{\sqrt{6}}}+{{2\,{ x_1}\,
 { y_1}}\over{\sqrt{15}}} \right) $$
$$ {\bf 4} \otimes {\bf 6} : {\bf 4'} $$ 
$$\left( {{{ x_1}\,{ y_6}}\over{\sqrt{30}}}+{{{ x_3}\,
 { y_5}}\over{\sqrt{6}}}+{{{ x_2}\,{ y_4}}\over{\sqrt{3}}}+
 {{{ x_4}\,{ y_3}}\over{\sqrt{3}}}+{{2\,{ x_1}\,{ y_1}
 }\over{\sqrt{30}}} , {{2\,{ x_3}\,{ y_6}}\over{\sqrt{10}}}-{{
 { x_1}\,{ y_2}}\over{\sqrt{2}}}-{{{ x_3}\,{ y_1}}\over{
 \sqrt{10}}} , 
  \right.
 $$
 $$
 \left.
 -{{{ x_2}\,{ y_6}}\over{\sqrt{10}}}-{{{ x_4}
 \,{ y_5}}\over{\sqrt{2}}}-{{2\,{ x_2}\,{ y_1}}\over{\sqrt{10
 }}} , {{2\,{ x_4}\,{ y_6}}\over{\sqrt{30}}}+{{{ x_1}\,
 { y_4}}\over{\sqrt{3}}}+{{{ x_3}\,{ y_3}}\over{\sqrt{3}}}-
 {{{ x_2}\,{ y_2}}\over{\sqrt{6}}}-{{{ x_4}\,{ y_1}
 }\over{\sqrt{30}}} \right) $$
$$ {\bf 4} \otimes {\bf 6} : {\bf 2} $$ 
$$\left( {{3\,{ x_3}\,{ y_6}}\over{\sqrt{30}}}+{{{ x_2}\,
 { y_5}}\over{\sqrt{6}}}+{{{ x_4}\,{ y_4}}\over{\sqrt{3}}}+
 {{{ x_1}\,{ y_2}}\over{\sqrt{6}}}+{{{ x_3}\,{ y_1}
 }\over{\sqrt{30}}} , -{{{ x_2}\,{ y_6}}\over{\sqrt{30}}}-{{
 { x_4}\,{ y_5}}\over{\sqrt{6}}}-{{{ x_1}\,{ y_3}}\over{
 \sqrt{3}}}-{{{ x_3}\,{ y_2}}\over{\sqrt{6}}}+{{3\,{ x_2}\,
 { y_1}}\over{\sqrt{30}}} \right) $$
$$ {\bf 4} \otimes {\bf 6} : {\bf 6} $$ 
$$\left( {{{ x_1}\,{ y_5}}\over{\sqrt{15}}}+{{2\,{ x_3}\,
 { y_4}}\over{\sqrt{30}}}+{{4\,{ x_2}\,{ y_3}}\over{\sqrt{30
 }}}+{{2\,{ x_4}\,{ y_2}}\over{\sqrt{15}}} , {{{ x_1}\,
 { y_6}}\over{\sqrt{15}}}+{{{ x_3}\,{ y_5}}\over{\sqrt{3}}}-
 {{3\,{ x_1}\,{ y_1}}\over{\sqrt{15}}} , {{3\,{ x_3}\,
 { y_6}}\over{\sqrt{30}}}-{{{ x_2}\,{ y_5}}\over{\sqrt{6}}}-
 {{{ x_4}\,{ y_4}}\over{\sqrt{3}}}+{{{ x_1}\,{ y_2}
 }\over{\sqrt{6}}}+{{{ x_3}\,{ y_1}}\over{\sqrt{30}}} , 
  \right.
 $$
 $$
 \left.
 -{{ { x_2}\,{ y_6}}\over{\sqrt{30}}}-{{{ x_4}\,{ y_5}}\over{
 \sqrt{6}}}+{{{ x_1}\,{ y_3}}\over{\sqrt{3}}}+{{{ x_3}\,
 { y_2}}\over{\sqrt{6}}}+{{3\,{ x_2}\,{ y_1}}\over{\sqrt{30}
 }} , {{3\,{ x_4}\,{ y_6}}\over{\sqrt{15}}}+{{{ x_2}\,
 { y_2}}\over{\sqrt{3}}}+{{{ x_4}\,{ y_1}}\over{\sqrt{15}}}
  , -{{2\,{ x_1}\,{ y_5}}\over{\sqrt{15}}}-{{4\,{ x_3}\,
 { y_4}}\over{\sqrt{30}}}+{{2\,{ x_2}\,{ y_3}}\over{\sqrt{30
 }}}+{{{ x_4}\,{ y_2}}\over{\sqrt{15}}} \right) $$
$$ {\bf 4} \otimes {\bf 6} : {\bf 4'} $$ 
$$\left( {{{ x_1}\,{ y_6}}\over{\sqrt{2}}}-{{{ x_3}\,
 { y_5}}\over{\sqrt{10}}}+{{{ x_2}\,{ y_4}}\over{\sqrt{5}}}-
 {{{ x_4}\,{ y_3}}\over{\sqrt{5}}} , -{{4\,{ x_2}\,{ y_5}
 }\over{\sqrt{30}}}+{{2\,{ x_4}\,{ y_4}}\over{\sqrt{15}}}-{{
 { x_1}\,{ y_2}}\over{\sqrt{30}}}+{{{ x_3}\,{ y_1}}\over{
 \sqrt{6}}} , 
  \right.
 $$
 $$
 \left.
 {{{ x_2}\,{ y_6}}\over{\sqrt{6}}}-{{{ x_4}\,
 { y_5}}\over{\sqrt{30}}}+{{2\,{ x_1}\,{ y_3}}\over{\sqrt{15
 }}}-{{4\,{ x_3}\,{ y_2}}\over{\sqrt{30}}} , -{{{ x_1}\,
 { y_4}}\over{\sqrt{5}}}+{{{ x_3}\,{ y_3}}\over{\sqrt{5}}}+
 {{{ x_2}\,{ y_2}}\over{\sqrt{10}}}-{{{ x_4}\,{ y_1}
 }\over{\sqrt{2}}} \right) $$
$$ {\bf 4} \otimes {\bf 6} : {\bf 6} $$ 
$$\left( {{{ x_1}\,{ y_5}}\over{\sqrt{3}}}-{{{ x_3}\,
 { y_4}}\over{\sqrt{6}}}+{{{ x_2}\,{ y_3}}\over{\sqrt{6}}}-
 {{{ x_4}\,{ y_2}}\over{\sqrt{3}}} , {{{ x_1}\,{ y_6}
 }\over{\sqrt{3}}}-{{{ x_3}\,{ y_5}}\over{\sqrt{15}}}-{{3\,
 { x_2}\,{ y_4}}\over{\sqrt{30}}}+{{3\,{ x_4}\,{ y_3}
 }\over{\sqrt{30}}} , -{{2\,{ x_2}\,{ y_5}}\over{\sqrt{30}}}+{{
 { x_4}\,{ y_4}}\over{\sqrt{15}}}+{{2\,{ x_1}\,{ y_2}
 }\over{\sqrt{30}}}-{{2\,{ x_3}\,{ y_1}}\over{\sqrt{6}}} , 
  \right.
 $$
 $$
 \left.
 {{2 \,{ x_2}\,{ y_6}}\over{\sqrt{6}}}-{{2\,{ x_4}\,{ y_5}
 }\over{\sqrt{30}}}-{{{ x_1}\,{ y_3}}\over{\sqrt{15}}}+{{2\,
 { x_3}\,{ y_2}}\over{\sqrt{30}}} , -{{3\,{ x_1}\,{ y_4}
 }\over{\sqrt{30}}}+{{3\,{ x_3}\,{ y_3}}\over{\sqrt{30}}}-{{
 { x_2}\,{ y_2}}\over{\sqrt{15}}}+{{{ x_4}\,{ y_1}}\over{
 \sqrt{3}}} , {{{ x_1}\,{ y_5}}\over{\sqrt{3}}}-{{{ x_3}\,
 { y_4}}\over{\sqrt{6}}}-{{{ x_2}\,{ y_3}}\over{\sqrt{6}}}+
 {{{ x_4}\,{ y_2}}\over{\sqrt{3}}} \right) $$

$\bf4'\otimes4'=(1)\oplus(3)\oplus(5)\oplus(3'\oplus4)$ 
$$ {\bf 4'} \otimes {\bf 4'} : {\bf 1} $$ 
$$\left( { x_1}\,{ y_4}-{ x_2}\,{ y_3}+{ x_3}\,
 { y_2}-{ x_4}\,{ y_1} \right) $$
$$ {\bf 4'} \otimes {\bf 4'} : {\bf 3} $$ 
$$\left( {{{ x_1}\,{ y_3}}\over{\sqrt{3}}}-{{2\,{ x_2}\,
 { y_2}}\over{3}}+{{{ x_3}\,{ y_1}}\over{\sqrt{3}}} , {{
 { x_1}\,{ y_4}}\over{\sqrt{2}}}-{{{ x_2}\,{ y_3}}\over{3
 \,\sqrt{2}}}-{{{ x_3}\,{ y_2}}\over{3\,\sqrt{2}}}+{{{ x_4}
 \,{ y_1}}\over{\sqrt{2}}} , {{{ x_2}\,{ y_4}}\over{\sqrt{3}
 }}-{{2\,{ x_3}\,{ y_3}}\over{3}}+{{{ x_4}\,{ y_2}}\over{
 \sqrt{3}}} \right) $$
$$ {\bf 4'} \otimes {\bf 4'} : {\bf 5} $$ 
$$\left( {{{ x_1}\,{ y_2}}\over{\sqrt{2}}}-{{{ x_2}\,
 { y_1}}\over{\sqrt{2}}} , {{{ x_1}\,{ y_3}}\over{\sqrt{2}}}
 -{{{ x_3}\,{ y_1}}\over{\sqrt{2}}} , {{{ x_1}\,{ y_4}
 }\over{2}}+{{{ x_2}\,{ y_3}}\over{2}}-{{{ x_3}\,{ y_2}
 }\over{2}}-{{{ x_4}\,{ y_1}}\over{2}} , {{{ x_2}\,{ y_4}
 }\over{\sqrt{2}}}-{{{ x_4}\,{ y_2}}\over{\sqrt{2}}} , {{
 { x_3}\,{ y_4}}\over{\sqrt{2}}}-{{{ x_4}\,{ y_3}}\over{
 \sqrt{2}}} \right) $$
$$ {\bf 4'} \otimes {\bf 4'} : {\bf 3'} $$ 
$$\left( {{4\,{ x_4}\,{ y_4}}\over{\sqrt{2}}}-{{6\,{ x_1}\,
 { y_2}}\over{\sqrt{6}}}-{{6\,{ x_2}\,{ y_1}}\over{\sqrt{6}
 }} , { x_1}\,{ y_4}+3\,{ x_2}\,{ y_3}+3\,{ x_3}\,
 { y_2}+{ x_4}\,{ y_1} , {{6\,{ x_3}\,{ y_4}}\over{
 \sqrt{6}}}+{{6\,{ x_4}\,{ y_3}}\over{\sqrt{6}}}+{{4\,{ x_1}
 \,{ y_1}}\over{\sqrt{2}}} \right) $$
$$ {\bf 4'} \otimes {\bf 4'} : {\bf 4} $$ 
$$\left( { x_2}\,{ y_4}+{{3\,{ x_3}\,{ y_3}}\over{\sqrt{3
 }}}+{ x_4}\,{ y_2} , -{{3\,{ x_4}\,{ y_4}}\over{\sqrt{3}
 }}-{ x_1}\,{ y_2}-{ x_2}\,{ y_1} , -{ x_3}\,{ y_4}
 -{ x_4}\,{ y_3}+{{3\,{ x_1}\,{ y_1}}\over{\sqrt{3}}} , 
 { x_1}\,{ y_3}+{{3\,{ x_2}\,{ y_2}}\over{\sqrt{3}}}+
 { x_3}\,{ y_1} \right) $$
 
$\bf4'\otimes5=(2)\oplus(4')\oplus(6)\oplus(2'\oplus6)$ 
$$ {\bf 4'} \otimes {\bf 5} : {\bf 2} $$ 
$$\left( {{{ x_1}\,{ y_4}}\over{2}}-{{{ x_2}\,{ y_3}
 }\over{\sqrt{2}}}+{{3\,{ x_3}\,{ y_2}}\over{2\,\sqrt{3}}}-
 { x_4}\,{ y_1} , { x_1}\,{ y_5}-{{3\,{ x_2}\,
 { y_4}}\over{2\,\sqrt{3}}}+{{{ x_3}\,{ y_3}}\over{\sqrt{2}
 }}-{{{ x_4}\,{ y_2}}\over{2}} \right) $$
$$ {\bf 4'} \otimes {\bf 5} : {\bf 4'} $$ 
$$\left( {{{ x_1}\,{ y_3}}\over{\sqrt{2}}}-{ x_2}\,{ y_2}
 +{ x_3}\,{ y_1} , { x_1}\,{ y_4}-{{{ x_2}\,{ y_3}
 }\over{\sqrt{2}}}+{ x_4}\,{ y_1} , { x_1}\,{ y_5}-{{
 { x_3}\,{ y_3}}\over{\sqrt{2}}}+{ x_4}\,{ y_2} , 
 { x_2}\,{ y_5}-{ x_3}\,{ y_4}+{{{ x_4}\,{ y_3}
 }\over{\sqrt{2}}} \right) $$

$$ {\bf 4'} \otimes {\bf 5} : {\bf 6} $$ 
$$\left( {{{ x_1}\,{ y_2}}\over{2}}-{{{ x_2}\,{ y_1}
 }\over{\sqrt{3}}} , {{3\,{ x_1}\,{ y_3}}\over{\sqrt{30}}}-{{
 { x_2}\,{ y_2}}\over{2\,\sqrt{15}}}-{{2\,{ x_3}\,{ y_1}
 }\over{\sqrt{15}}} , {{3\,{ x_1}\,{ y_4}}\over{2\,\sqrt{10}}}+
 {{{ x_2}\,{ y_3}}\over{2\,\sqrt{5}}}-{{\sqrt{5}\,{ x_3}\,
 { y_2}}\over{2\,\sqrt{6}}}-{{{ x_4}\,{ y_1}}\over{\sqrt{10}
 }} , 
  \right.
 $$
 $$
 \left.
 {{{ x_1}\,{ y_5}}\over{\sqrt{10}}}+{{\sqrt{5}\,{ x_2}
 \,{ y_4}}\over{2\,\sqrt{6}}}-{{{ x_3}\,{ y_3}}\over{2\,
 \sqrt{5}}}-{{3\,{ x_4}\,{ y_2}}\over{2\,\sqrt{10}}} , {{2\,
 { x_2}\,{ y_5}}\over{\sqrt{15}}}+{{{ x_3}\,{ y_4}}\over{2
 \,\sqrt{15}}}-{{3\,{ x_4}\,{ y_3}}\over{\sqrt{30}}} , {{
 { x_3}\,{ y_5}}\over{\sqrt{3}}}-{{{ x_4}\,{ y_4}}\over{2
 }} \right) $$
$$ {\bf 4'} \otimes {\bf 5} : {\bf 2'} $$ 
$$\left( -{ x_2}\,{ y_5}-2\,{ x_3}\,{ y_4}-{{2\,{ x_4}
 \,{ y_3}}\over{\sqrt{2}}}+{{3\,{ x_1}\,{ y_1}}\over{\sqrt{3
 }}} , {{3\,{ x_4}\,{ y_5}}\over{\sqrt{3}}}+{{2\,{ x_1}\,
 { y_3}}\over{\sqrt{2}}}+2\,{ x_2}\,{ y_2}+{ x_3}\,
 { y_1} \right) $$
$$ {\bf 4'} \otimes {\bf 5} : {\bf 6} $$ 
$$\left( -{{7\,{ x_3}\,{ y_5}}\over{\sqrt{5}}}-{{14\,{ x_4}
 \,{ y_4}}\over{\sqrt{15}}}+{{2\,{ x_1}\,{ y_2}}\over{\sqrt{15
 }}}+{{{ x_2}\,{ y_1}}\over{\sqrt{5}}} , {{7\,{ x_4}\,
 { y_5}}\over{\sqrt{3}}}-{{2\,{ x_1}\,{ y_3}}\over{\sqrt{2}
 }}-2\,{ x_2}\,{ y_2}-{ x_3}\,{ y_1} , 
  \right.
 $$
 $$
 \left.
 {{4\,{ x_1}\,
 { y_4}}\over{\sqrt{6}}}+{{6\,{ x_2}\,{ y_3}}\over{\sqrt{3}
 }}+{{4\,{ x_3}\,{ y_2}}\over{\sqrt{2}}}+{{2\,{ x_4}\,
 { y_1}}\over{\sqrt{6}}} , -{{2\,{ x_1}\,{ y_5}}\over{\sqrt{6
 }}}-{{4\,{ x_2}\,{ y_4}}\over{\sqrt{2}}}-{{6\,{ x_3}\,
 { y_3}}\over{\sqrt{3}}}-{{4\,{ x_4}\,{ y_2}}\over{\sqrt{6}
 }} , 
  \right.
 $$
 $$
 \left.
 { x_2}\,{ y_5}+2\,{ x_3}\,{ y_4}+{{2\,{ x_4}\,
 { y_3}}\over{\sqrt{2}}}+{{7\,{ x_1}\,{ y_1}}\over{\sqrt{3}
 }} , -{{{ x_3}\,{ y_5}}\over{\sqrt{5}}}-{{2\,{ x_4}\,
 { y_4}}\over{\sqrt{15}}}-{{14\,{ x_1}\,{ y_2}}\over{\sqrt{15
 }}}-{{7\,{ x_2}\,{ y_1}}\over{\sqrt{5}}} \right) $$

$\bf4'\otimes6=(3)\oplus(5)\oplus(3'\oplus4)\oplus(4\oplus5)$ 
$$ {\bf 4'} \otimes {\bf 6} : {\bf 3} $$ 
$$\left( {{{ x_1}\,{ y_4}}\over{2\,\sqrt{5}}}-{{3\,{ x_2}\,
 { y_3}}\over{2\,\sqrt{15}}}+{{3\,{ x_3}\,{ y_2}}\over{
 \sqrt{30}}}-{{{ x_4}\,{ y_1}}\over{\sqrt{2}}} , {{{ x_1}\,
 { y_5}}\over{\sqrt{5}}}-{{3\,{ x_2}\,{ y_4}}\over{\sqrt{30}
 }}+{{3\,{ x_3}\,{ y_3}}\over{\sqrt{30}}}-{{{ x_4}\,
 { y_2}}\over{\sqrt{5}}} , {{{ x_1}\,{ y_6}}\over{\sqrt{2}}}
 -{{3\,{ x_2}\,{ y_5}}\over{\sqrt{30}}}+{{3\,{ x_3}\,
 { y_4}}\over{2\,\sqrt{15}}}-{{{ x_4}\,{ y_3}}\over{2\,
 \sqrt{5}}} \right) $$

$$ {\bf 4'} \otimes {\bf 6} : {\bf 5} $$ 
$$\left( {{{ x_1}\,{ y_3}}\over{\sqrt{10}}}-{{2\,{ x_2}\,
 { y_2}}\over{\sqrt{15}}}+{{{ x_3}\,{ y_1}}\over{\sqrt{3}}}
  , {{3\,{ x_1}\,{ y_4}}\over{2\,\sqrt{10}}}-{{\sqrt{5}\,
 { x_2}\,{ y_3}}\over{2\,\sqrt{6}}}+{{{ x_3}\,{ y_2}
 }\over{2\,\sqrt{15}}}+{{{ x_4}\,{ y_1}}\over{2}} , {{3\,
 { x_1}\,{ y_5}}\over{\sqrt{30}}}-{{{ x_2}\,{ y_4}}\over{2
 \,\sqrt{5}}}-{{{ x_3}\,{ y_3}}\over{2\,\sqrt{5}}}+{{3\,
 { x_4}\,{ y_2}}\over{\sqrt{30}}} , 
  \right.
 $$
 $$
 \left.
 {{{ x_1}\,{ y_6}
 }\over{2}}+{{{ x_2}\,{ y_5}}\over{2\,\sqrt{15}}}-{{\sqrt{5}\,
 { x_3}\,{ y_4}}\over{2\,\sqrt{6}}}+{{3\,{ x_4}\,{ y_3}
 }\over{2\,\sqrt{10}}} , {{{ x_2}\,{ y_6}}\over{\sqrt{3}}}-{{2
 \,{ x_3}\,{ y_5}}\over{\sqrt{15}}}+{{{ x_4}\,{ y_4}
 }\over{\sqrt{10}}} \right) $$
$$ {\bf 4'} \otimes {\bf 6} : {\bf 3'} $$ 
$$\left( {{{ x_3}\,{ y_6}}\over{2}}-{{3\,{ x_4}\,{ y_5}
 }\over{2\,\sqrt{15}}}-{{3\,{ x_1}\,{ y_3}}\over{\sqrt{30}}}+{{
 { x_2}\,{ y_2}}\over{2\,\sqrt{5}}}+{{{ x_3}\,{ y_1}
 }\over{2}} , {{3\,{ x_1}\,{ y_5}}\over{\sqrt{30}}}+{{{ x_2}
 \,{ y_4}}\over{\sqrt{5}}}-{{{ x_3}\,{ y_3}}\over{\sqrt{5}}}
 -{{3\,{ x_4}\,{ y_2}}\over{\sqrt{30}}} , 
  \right.
 $$
 $$
 \left.
 {{{ x_2}\,
 { y_6}}\over{2}}+{{{ x_3}\,{ y_5}}\over{2\,\sqrt{5}}}-{{3\,
 { x_4}\,{ y_4}}\over{\sqrt{30}}}+{{3\,{ x_1}\,{ y_2}
 }\over{2\,\sqrt{15}}}-{{{ x_2}\,{ y_1}}\over{2}} \right) $$

$$ {\bf 4'} \otimes {\bf 6} : {\bf 4} $$ 
$$\left( {{{ x_1}\,{ y_6}}\over{2\,\sqrt{2}}}+{{7\,{ x_2}\,
 { y_5}}\over{2\,\sqrt{30}}}-{{{ x_3}\,{ y_4}}\over{2\,
 \sqrt{15}}}-{{3\,{ x_4}\,{ y_3}}\over{2\,\sqrt{5}}} , -{{3\,
 { x_3}\,{ y_6}}\over{2\,\sqrt{6}}}+{{3\,{ x_4}\,{ y_5}
 }\over{2\,\sqrt{10}}}-{{{ x_1}\,{ y_3}}\over{\sqrt{5}}}+{{
 { x_2}\,{ y_2}}\over{\sqrt{30}}}+{{{ x_3}\,{ y_1}}\over{
 \sqrt{6}}} , 
  \right.
 $$
 $$
 \left.
 -{{{ x_2}\,{ y_6}}\over{\sqrt{6}}}-{{{ x_3}\,
 { y_5}}\over{\sqrt{30}}}+{{{ x_4}\,{ y_4}}\over{\sqrt{5}}}+
 {{3\,{ x_1}\,{ y_2}}\over{2\,\sqrt{10}}}-{{3\,{ x_2}\,
 { y_1}}\over{2\,\sqrt{6}}} , {{3\,{ x_1}\,{ y_4}}\over{2\,
 \sqrt{5}}}+{{{ x_2}\,{ y_3}}\over{2\,\sqrt{15}}}-{{7\,
 { x_3}\,{ y_2}}\over{2\,\sqrt{30}}}-{{{ x_4}\,{ y_1}
 }\over{2\,\sqrt{2}}} \right) $$

$$ {\bf 4'} \otimes {\bf 6} : {\bf 4} $$ 
$$\left( -{{{ x_1}\,{ y_6}}\over{2\,\sqrt{30}}}-{{{ x_2}\,
 { y_5}}\over{2\,\sqrt{2}}}-{{{ x_3}\,{ y_4}}\over{2}}-{{
 { x_4}\,{ y_3}}\over{2\,\sqrt{3}}}+{{4\,{ x_1}\,{ y_1}
 }\over{\sqrt{30}}} , {{{ x_3}\,{ y_6}}\over{2\,\sqrt{10}}}+{{
 { x_4}\,{ y_5}}\over{2\,\sqrt{6}}}+{{{ x_1}\,{ y_3}
 }\over{\sqrt{3}}}+{{{ x_2}\,{ y_2}}\over{\sqrt{2}}}+{{
 { x_3}\,{ y_1}}\over{\sqrt{10}}} , 
  \right.
 $$
 $$
 \left.
 -{{{ x_2}\,{ y_6}
 }\over{\sqrt{10}}}-{{{ x_3}\,{ y_5}}\over{\sqrt{2}}}-{{
 { x_4}\,{ y_4}}\over{\sqrt{3}}}+{{{ x_1}\,{ y_2}}\over{2
 \,\sqrt{6}}}+{{{ x_2}\,{ y_1}}\over{2\,\sqrt{10}}} , {{4\,
 { x_4}\,{ y_6}}\over{\sqrt{30}}}+{{{ x_1}\,{ y_4}}\over{2
 \,\sqrt{3}}}+{{{ x_2}\,{ y_3}}\over{2}}+{{{ x_3}\,{ y_2}
 }\over{2\,\sqrt{2}}}+{{{ x_4}\,{ y_1}}\over{2\,\sqrt{30}}}
  \right) $$
$$ {\bf 4'} \otimes {\bf 6} : {\bf 5} $$ 
$$\left( -{{7\,{ x_3}\,{ y_6}}\over{2\,\sqrt{5}}}-{{7\,
 { x_4}\,{ y_5}}\over{2\,\sqrt{3}}}+{{{ x_1}\,{ y_3}
 }\over{\sqrt{6}}}+{{{ x_2}\,{ y_2}}\over{2}}+{{{ x_3}\,
 { y_1}}\over{2\,\sqrt{5}}} , {{7\,{ x_4}\,{ y_6}}\over{
 \sqrt{15}}}-{{2\,{ x_1}\,{ y_4}}\over{\sqrt{6}}}-{{2\,
 { x_2}\,{ y_3}}\over{\sqrt{2}}}-{ x_3}\,{ y_2}-{{
 { x_4}\,{ y_1}}\over{\sqrt{15}}} , 
  \right.
 $$
 $$
 \left.
 {{{ x_1}\,{ y_5}
 }\over{\sqrt{2}}}+{{3\,{ x_2}\,{ y_4}}\over{\sqrt{3}}}+{{3\,
 { x_3}\,{ y_3}}\over{\sqrt{3}}}+{{{ x_4}\,{ y_2}}\over{
 \sqrt{2}}} , 
  \right.
 $$
 $$
 \left.
 -{{{ x_1}\,{ y_6}}\over{\sqrt{15}}}-{ x_2}\,
 { y_5}-{{2\,{ x_3}\,{ y_4}}\over{\sqrt{2}}}-{{2\,{ x_4}
 \,{ y_3}}\over{\sqrt{6}}}-{{7\,{ x_1}\,{ y_1}}\over{\sqrt{15
 }}} , {{{ x_2}\,{ y_6}}\over{2\,\sqrt{5}}}+{{{ x_3}\,
 { y_5}}\over{2}}+{{{ x_4}\,{ y_4}}\over{\sqrt{6}}}+{{7\,
 { x_1}\,{ y_2}}\over{2\,\sqrt{3}}}+{{7\,{ x_2}\,{ y_1}
 }\over{2\,\sqrt{5}}} \right) $$

$\bf5\otimes5=(1)\oplus(3)\oplus(5)\oplus(3'\oplus4)\oplus(4\oplus5)$ 
$$ {\bf 5} \otimes {\bf 5} : {\bf 1} $$ 
$$\left( { x_1}\,{ y_5}-{ x_2}\,{ y_4}+{ x_3}\,
 { y_3}-{ x_4}\,{ y_2}+{ x_5}\,{ y_1} \right) $$
$$ {\bf 5} \otimes {\bf 5} : {\bf 3} $$ 
$$\left( {{{ x_1}\,{ y_4}}\over{\sqrt{5}}}-{{3\,{ x_2}\,
 { y_3}}\over{\sqrt{30}}}+{{3\,{ x_3}\,{ y_2}}\over{\sqrt{30
 }}}-{{{ x_4}\,{ y_1}}\over{\sqrt{5}}} , {{2\,{ x_1}\,
 { y_5}}\over{\sqrt{10}}}-{{{ x_2}\,{ y_4}}\over{\sqrt{10}}}
 +{{{ x_4}\,{ y_2}}\over{\sqrt{10}}}-{{2\,{ x_5}\,{ y_1}
 }\over{\sqrt{10}}} , {{{ x_2}\,{ y_5}}\over{\sqrt{5}}}-{{3\,
 { x_3}\,{ y_4}}\over{\sqrt{30}}}+{{3\,{ x_4}\,{ y_3}
 }\over{\sqrt{30}}}-{{{ x_5}\,{ y_2}}\over{\sqrt{5}}} \right) $$
$$ {\bf 5} \otimes {\bf 5} : {\bf 5} $$ 
$$\left( { x_1}\,{ y_3}-{{3\,{ x_2}\,{ y_2}}\over{\sqrt{6
 }}}+{ x_3}\,{ y_1} , {{3\,{ x_1}\,{ y_4}}\over{\sqrt{6}
 }}-{{{ x_2}\,{ y_3}}\over{2}}-{{{ x_3}\,{ y_2}}\over{2}}
 +{{3\,{ x_4}\,{ y_1}}\over{\sqrt{6}}} , { x_1}\,{ y_5}+
 {{{ x_2}\,{ y_4}}\over{2}}-{ x_3}\,{ y_3}+{{{ x_4}\,
 { y_2}}\over{2}}+{ x_5}\,{ y_1} , 
  \right.
 $$
 $$
 \left.
 {{3\,{ x_2}\,{ y_5}
 }\over{\sqrt{6}}}-{{{ x_3}\,{ y_4}}\over{2}}-{{{ x_4}\,
 { y_3}}\over{2}}+{{3\,{ x_5}\,{ y_2}}\over{\sqrt{6}}} , 
 { x_3}\,{ y_5}-{{3\,{ x_4}\,{ y_4}}\over{\sqrt{6}}}+
 { x_5}\,{ y_3} \right) $$
$$ {\bf 5} \otimes {\bf 5} : {\bf 3'} $$ 
$$\left( {{2\,{ x_4}\,{ y_5}}\over{\sqrt{2}}}-{{2\,{ x_5}\,
 { y_4}}\over{\sqrt{2}}}-{{3\,{ x_1}\,{ y_3}}\over{\sqrt{3}
 }}+{{3\,{ x_3}\,{ y_1}}\over{\sqrt{3}}} , { x_1}\,{ y_5}
 +2\,{ x_2}\,{ y_4}-2\,{ x_4}\,{ y_2}-{ x_5}\,
 { y_1} ,
  \right.
 $$
 $$
 \left.
  {{3\,{ x_3}\,{ y_5}}\over{\sqrt{3}}}-{{3\,
 { x_5}\,{ y_3}}\over{\sqrt{3}}}+{{2\,{ x_1}\,{ y_2}
 }\over{\sqrt{2}}}-{{2\,{ x_2}\,{ y_1}}\over{\sqrt{2}}}
  \right) $$
$$ {\bf 5} \otimes {\bf 5} : {\bf 4} $$ 
$$\left( {{3\,{ x_2}\,{ y_5}}\over{\sqrt{6}}}+{ x_3}\,
 { y_4}-{ x_4}\,{ y_3}-{{3\,{ x_5}\,{ y_2}}\over{
 \sqrt{6}}} , -{{3\,{ x_4}\,{ y_5}}\over{\sqrt{6}}}+{{3\,
 { x_5}\,{ y_4}}\over{\sqrt{6}}}-{ x_1}\,{ y_3}+{ x_3}
 \,{ y_1} , 
  \right.
 $$
 $$
 \left.
 -{ x_3}\,{ y_5}+{ x_5}\,{ y_3}+{{3\,
 { x_1}\,{ y_2}}\over{\sqrt{6}}}-{{3\,{ x_2}\,{ y_1}
 }\over{\sqrt{6}}} , {{3\,{ x_1}\,{ y_4}}\over{\sqrt{6}}}+
 { x_2}\,{ y_3}-{ x_3}\,{ y_2}-{{3\,{ x_4}\,{ y_1}
 }\over{\sqrt{6}}} \right) $$
$$ {\bf 5} \otimes {\bf 5} : {\bf 4} $$ 
$$\left( -{{{ x_2}\,{ y_5}}\over{\sqrt{30}}}-{{{ x_3}\,
 { y_4}}\over{\sqrt{5}}}-{{{ x_4}\,{ y_3}}\over{\sqrt{5}}}-
 {{{ x_5}\,{ y_2}}\over{\sqrt{30}}}+{{4\,{ x_1}\,{ y_1}
 }\over{\sqrt{30}}} , {{{ x_4}\,{ y_5}}\over{\sqrt{30}}}+{{
 { x_5}\,{ y_4}}\over{\sqrt{30}}}+{{{ x_1}\,{ y_3}}\over{
 \sqrt{5}}}+{{4\,{ x_2}\,{ y_2}}\over{\sqrt{30}}}+{{{ x_3}\,
 { y_1}}\over{\sqrt{5}}} , 
  \right.
 $$
 $$
 \left.
 -{{{ x_3}\,{ y_5}}\over{\sqrt{5}
 }}-{{4\,{ x_4}\,{ y_4}}\over{\sqrt{30}}}-{{{ x_5}\,
 { y_3}}\over{\sqrt{5}}}+{{{ x_1}\,{ y_2}}\over{\sqrt{30}}}+
 {{{ x_2}\,{ y_1}}\over{\sqrt{30}}} , {{4\,{ x_5}\,{ y_5}
 }\over{\sqrt{30}}}+{{{ x_1}\,{ y_4}}\over{\sqrt{30}}}+{{
 { x_2}\,{ y_3}}\over{\sqrt{5}}}+{{{ x_3}\,{ y_2}}\over{
 \sqrt{5}}}+{{{ x_4}\,{ y_1}}\over{\sqrt{30}}} \right) $$

$$ {\bf 5} \otimes {\bf 5} : {\bf 5} $$ 
$$\left( -{{7\,\sqrt{2}\,{ x_4}\,{ y_5}}\over{\sqrt{3}}}-{{7\,
 \sqrt{2}\,{ x_5}\,{ y_4}}\over{\sqrt{3}}}+{ x_1}\,{ y_3}
 +{{4\,{ x_2}\,{ y_2}}\over{\sqrt{6}}}+{ x_3}\,{ y_1} , 
 {{7\,\sqrt{2}\,{ x_5}\,{ y_5}}\over{\sqrt{3}}}-{{4\,{ x_1}
 \,{ y_4}}\over{\sqrt{6}}}-4\,{ x_2}\,{ y_3}-4\,{ x_3}\,
 { y_2}-{{4\,{ x_4}\,{ y_1}}\over{\sqrt{6}}} , 
  \right.
 $$
 $$
 \left.
 { x_1}\,
 { y_5}+4\,{ x_2}\,{ y_4}+6\,{ x_3}\,{ y_3}+4\,
 { x_4}\,{ y_2}+{ x_5}\,{ y_1} , 
  \right.
 $$
 $$
 \left.
 -{{4\,{ x_2}\,
 { y_5}}\over{\sqrt{6}}}-4\,{ x_3}\,{ y_4}-4\,{ x_4}\,
 { y_3}-{{4\,{ x_5}\,{ y_2}}\over{\sqrt{6}}}-{{7\,\sqrt{2}\,
 { x_1}\,{ y_1}}\over{\sqrt{3}}} , 
 { x_3}\,{ y_5}+{{4\,
 { x_4}\,{ y_4}}\over{\sqrt{6}}}+{ x_5}\,{ y_3}+{{7\,
 \sqrt{2}\,{ x_1}\,{ y_2}}\over{\sqrt{3}}}+{{7\,\sqrt{2}\,
 { x_2}\,{ y_1}}\over{\sqrt{3}}} \right) $$

$\bf5\otimes6=(2)\oplus(4')\oplus(6)\oplus(2'\oplus6)\oplus(4'\oplus6)$ 
$$ {\bf 5} \otimes {\bf 6} : {\bf 2} $$ 
$$\left( {{{ x_1}\,{ y_5}}\over{\sqrt{5}}}-{{2\,{ x_2}\,
 { y_4}}\over{\sqrt{10}}}+{{3\,{ x_3}\,{ y_3}}\over{\sqrt{15
 }}}-{{2\,{ x_4}\,{ y_2}}\over{\sqrt{5}}}+{ x_5}\,{ y_1}
  , { x_1}\,{ y_6}-{{2\,{ x_2}\,{ y_5}}\over{\sqrt{5}}}+
 {{3\,{ x_3}\,{ y_4}}\over{\sqrt{15}}}-{{2\,{ x_4}\,
 { y_3}}\over{\sqrt{10}}}+{{{ x_5}\,{ y_2}}\over{\sqrt{5}}}
  \right) $$

$$ {\bf 5} \otimes {\bf 6} : {\bf 4'} $$ 
$$\left( {{{ x_1}\,{ y_4}}\over{\sqrt{10}}}-{{3\,{ x_2}\,
 { y_3}}\over{2\,\sqrt{10}}}+{{3\,{ x_3}\,{ y_2}}\over{
 \sqrt{30}}}-{{{ x_4}\,{ y_1}}\over{2}} , {{2\,{ x_1}\,
 { y_5}}\over{\sqrt{15}}}-{{\sqrt{5}\,{ x_2}\,{ y_4}}\over{2
 \,\sqrt{6}}}+{{{ x_3}\,{ y_3}}\over{2\,\sqrt{5}}}+{{{ x_4}
 \,{ y_2}}\over{2\,\sqrt{15}}}-{{{ x_5}\,{ y_1}}\over{\sqrt{3
 }}} , 
  \right.
 $$
 $$
 \left.
 {{{ x_1}\,{ y_6}}\over{\sqrt{3}}}-{{{ x_2}\,{ y_5}
 }\over{2\,\sqrt{15}}}-{{{ x_3}\,{ y_4}}\over{2\,\sqrt{5}}}+{{
 \sqrt{5}\,{ x_4}\,{ y_3}}\over{2\,\sqrt{6}}}-{{2\,{ x_5}\,
 { y_2}}\over{\sqrt{15}}} , {{{ x_2}\,{ y_6}}\over{2}}-{{3\,
 { x_3}\,{ y_5}}\over{\sqrt{30}}}+{{3\,{ x_4}\,{ y_4}
 }\over{2\,\sqrt{10}}}-{{{ x_5}\,{ y_3}}\over{\sqrt{10}}}
  \right) $$
$$ {\bf 5} \otimes {\bf 6} : {\bf 6} $$ 
$$\left( {{{ x_1}\,{ y_3}}\over{\sqrt{2}}}-{ x_2}\,{ y_2}
 +{{5\,{ x_3}\,{ y_1}}\over{\sqrt{30}}} , 
 {{3\,{ x_1}\,
 { y_4}}\over{\sqrt{10}}}-{{2\,{ x_2}\,{ y_3}}\over{\sqrt{10
 }}}-{{{ x_3}\,{ y_2}}\over{\sqrt{30}}}+{ x_4}\,{ y_1} , 
 {{3\,{ x_1}\,{ y_5}}\over{\sqrt{10}}}-{{4\,{ x_3}\,
 { y_3}}\over{\sqrt{30}}}+{{2\,{ x_4}\,{ y_2}}\over{\sqrt{10
 }}}+{{{ x_5}\,{ y_1}}\over{\sqrt{2}}} , 
  \right.
 $$
 $$
 \left.
 {{{ x_1}\,{ y_6}
 }\over{\sqrt{2}}}+{{2\,{ x_2}\,{ y_5}}\over{\sqrt{10}}}-{{4\,
 { x_3}\,{ y_4}}\over{\sqrt{30}}}+{{3\,{ x_5}\,{ y_2}
 }\over{\sqrt{10}}} , { x_2}\,{ y_6}-{{{ x_3}\,{ y_5}
 }\over{\sqrt{30}}}-{{2\,{ x_4}\,{ y_4}}\over{\sqrt{10}}}+{{3\,
 { x_5}\,{ y_3}}\over{\sqrt{10}}} , {{5\,{ x_3}\,{ y_6}
 }\over{\sqrt{30}}}-{ x_4}\,{ y_5}+{{{ x_5}\,{ y_4}
 }\over{\sqrt{2}}} \right) $$

$$ {\bf 5} \otimes {\bf 6} : {\bf 6} $$ 
$$\left( -{{7\,\sqrt{3}\,{ x_3}\,{ y_6}}\over{2\,\sqrt{5}}}-{{7
 \,{ x_4}\,{ y_5}}\over{10\,\sqrt{2}}}+{{14\,{ x_5}\,
 { y_4}}\over{5}}+{{2\,{ x_1}\,{ y_3}}\over{5}}-{{{ x_2}
 \,{ y_2}}\over{10\,\sqrt{2}}}-{{3\,{ x_3}\,{ y_1}}\over{2\,
 \sqrt{15}}} , 
  \right.
 $$
 $$
 \left.
 {{7\,{ x_4}\,{ y_6}}\over{2\,\sqrt{2}}}-{{7\,
 { x_5}\,{ y_5}}\over{\sqrt{10}}}-{{3\,{ x_1}\,{ y_4}
 }\over{\sqrt{5}}}-{{3\,{ x_2}\,{ y_3}}\over{2\,\sqrt{5}}}+{{3
 \,\sqrt{3}\,{ x_3}\,{ y_2}}\over{2\,\sqrt{5}}}+{{3\,{ x_4}
 \,{ y_1}}\over{2\,\sqrt{2}}} , 
  \right.
 $$
 $$
 \left.
 {{4\,{ x_1}\,{ y_5}}\over{
 \sqrt{5}}}+{{7\,{ x_2}\,{ y_4}}\over{\sqrt{10}}}-{{3\,
 { x_3}\,{ y_3}}\over{\sqrt{15}}}-{{11\,{ x_4}\,{ y_2}
 }\over{2\,\sqrt{5}}}-{ x_5}\,{ y_1} , 
 -{ x_1}\,{ y_6}-{{11
 \,{ x_2}\,{ y_5}}\over{2\,\sqrt{5}}}-{{3\,{ x_3}\,{ y_4}
 }\over{\sqrt{15}}}+{{7\,{ x_4}\,{ y_3}}\over{\sqrt{10}}}+{{4\,
 { x_5}\,{ y_2}}\over{\sqrt{5}}} , 
   \right.
 $$
 $$
 \left.
 {{3\,{ x_2}\,{ y_6}
 }\over{2\,\sqrt{2}}}+{{3\,\sqrt{3}\,{ x_3}\,{ y_5}}\over{2\,
 \sqrt{5}}}-{{3\,{ x_4}\,{ y_4}}\over{2\,\sqrt{5}}}-{{3\,
 { x_5}\,{ y_3}}\over{\sqrt{5}}}+{{7\,{ x_1}\,{ y_2}
 }\over{\sqrt{10}}}-{{7\,{ x_2}\,{ y_1}}\over{2\,\sqrt{2}}} , 
   \right.
 $$
 $$
 \left.
 - {{3\,{ x_3}\,{ y_6}}\over{2\,\sqrt{15}}}-{{{ x_4}\,
 { y_5}}\over{10\,\sqrt{2}}}+{{2\,{ x_5}\,{ y_4}}\over{5}}-
 {{14\,{ x_1}\,{ y_3}}\over{5}}+{{7\,{ x_2}\,{ y_2}
 }\over{10\,\sqrt{2}}}+{{7\,\sqrt{3}\,{ x_3}\,{ y_1}}\over{2\,
 \sqrt{5}}} \right) $$

$$ {\bf 5} \otimes {\bf 6} : {\bf 2'} $$ 
$$\left( -{{{ x_2}\,{ y_6}}\over{\sqrt{6}}}-{{{ x_3}\,
 { y_5}}\over{\sqrt{5}}}+{{{ x_4}\,{ y_4}}\over{\sqrt{15}}}+
 {{2\,{ x_5}\,{ y_3}}\over{\sqrt{15}}}+{{2\,{ x_1}\,
 { y_2}}\over{\sqrt{30}}}-{{{ x_2}\,{ y_1}}\over{\sqrt{6}}}
  , {{{ x_4}\,{ y_6}}\over{\sqrt{6}}}-{{2\,{ x_5}\,{ y_5}
 }\over{\sqrt{30}}}+{{2\,{ x_1}\,{ y_4}}\over{\sqrt{15}}}+{{
 { x_2}\,{ y_3}}\over{\sqrt{15}}}-{{{ x_3}\,{ y_2}}\over{
 \sqrt{5}}}-{{{ x_4}\,{ y_1}}\over{\sqrt{6}}} \right) $$

$$ {\bf 5} \otimes {\bf 6} : {\bf 4'} $$ 
$$\left( {{14\,{ x_4}\,{ y_6}}\over{5\,\sqrt{3}}}+{{7\,
 { x_5}\,{ y_5}}\over{\sqrt{15}}}-{{2\,{ x_1}\,{ y_4}
 }\over{\sqrt{30}}}-{{4\,{ x_2}\,{ y_3}}\over{\sqrt{30}}}-{{2\,
 { x_3}\,{ y_2}}\over{\sqrt{10}}}-{{2\,{ x_4}\,{ y_1}
 }\over{5\,\sqrt{3}}} , 
   \right.
 $$
 $$
 \left.
 -{{7\,{ x_5}\,{ y_6}}\over{5}}+{{
 { x_1}\,{ y_5}}\over{\sqrt{5}}}+{{4\,{ x_2}\,{ y_4}
 }\over{\sqrt{10}}}+{{6\,{ x_3}\,{ y_3}}\over{\sqrt{15}}}+{{2\,
 { x_4}\,{ y_2}}\over{\sqrt{5}}}+{{{ x_5}\,{ y_1}}\over{5
 }} , 
   \right.
 $$
 $$
 \left.
 -{{{ x_1}\,{ y_6}}\over{5}}-{{2\,{ x_2}\,{ y_5}
 }\over{\sqrt{5}}}-{{6\,{ x_3}\,{ y_4}}\over{\sqrt{15}}}-{{4\,
 { x_4}\,{ y_3}}\over{\sqrt{10}}}-{{{ x_5}\,{ y_2}}\over{
 \sqrt{5}}}-{{7\,{ x_1}\,{ y_1}}\over{5}} , 
   \right.
 $$
 $$
 \left.
 {{2\,{ x_2}\,
 { y_6}}\over{5\,\sqrt{3}}}+{{2\,{ x_3}\,{ y_5}}\over{\sqrt{10
 }}}+{{4\,{ x_4}\,{ y_4}}\over{\sqrt{30}}}+{{2\,{ x_5}\,
 { y_3}}\over{\sqrt{30}}}+{{7\,{ x_1}\,{ y_2}}\over{\sqrt{15
 }}}+{{14\,{ x_2}\,{ y_1}}\over{5\,\sqrt{3}}} \right) $$

$$ {\bf 5} \otimes {\bf 6} : {\bf 6} $$ 
$$\left( -{{2\,{ x_3}\,{ y_6}}\over{\sqrt{30}}}-{{2\,{ x_4}
 \,{ y_5}}\over{3}}-{{2\,{ x_5}\,{ y_4}}\over{3\,\sqrt{2}}}+
 {{{ x_1}\,{ y_3}}\over{3\,\sqrt{2}}}+{{{ x_2}\,{ y_2}
 }\over{3}}+{{{ x_3}\,{ y_1}}\over{\sqrt{30}}} ,
   \right.
 $$
 $$
 \left.
  -{{4\,
 { x_4}\,{ y_6}}\over{15}}-{{2\,{ x_5}\,{ y_5}}\over{3\,
 \sqrt{5}}}-{{{ x_1}\,{ y_4}}\over{\sqrt{10}}}-{{2\,{ x_2}\,
 { y_3}}\over{\sqrt{10}}}-{{3\,{ x_3}\,{ y_2}}\over{\sqrt{30
 }}}-{{{ x_4}\,{ y_1}}\over{5}} , 
   \right.
 $$
 $$
 \left.
 {{3\,\sqrt{2}\,{ x_5}\,
 { y_6}}\over{5}}+{{{ x_1}\,{ y_5}}\over{3\,\sqrt{10}}}+{{2
 \,{ x_2}\,{ y_4}}\over{3\,\sqrt{5}}}+{{2\,{ x_3}\,{ y_3}
 }\over{\sqrt{30}}}+{{\sqrt{2}\,{ x_4}\,{ y_2}}\over{3\,\sqrt{5
 }}}+{{{ x_5}\,{ y_1}}\over{15\,\sqrt{2}}} , 
   \right.
 $$
 $$
 \left.
 {{{ x_1}\,
 { y_6}}\over{15\,\sqrt{2}}}+{{\sqrt{2}\,{ x_2}\,{ y_5}
 }\over{3\,\sqrt{5}}}+{{2\,{ x_3}\,{ y_4}}\over{\sqrt{30}}}+{{2
 \,{ x_4}\,{ y_3}}\over{3\,\sqrt{5}}}+{{{ x_5}\,{ y_2}
 }\over{3\,\sqrt{10}}}-{{3\,\sqrt{2}\,{ x_1}\,{ y_1}}\over{5}}, 
   \right.
 $$
 $$
 \left.
 -{{{ x_2}\,{ y_6}}\over{5}}-{{3\,{ x_3}\,{ y_5}
 }\over{\sqrt{30}}}-{{2\,{ x_4}\,{ y_4}}\over{\sqrt{10}}}-{{
 { x_5}\,{ y_3}}\over{\sqrt{10}}}+{{2\,{ x_1}\,{ y_2}
 }\over{3\,\sqrt{5}}}+{{4\,{ x_2}\,{ y_1}}\over{15}} , 
   \right.
 $$
 $$
 \left.
 {{
 { x_3}\,{ y_6}}\over{\sqrt{30}}}+{{{ x_4}\,{ y_5}}\over{3
 }}+{{{ x_5}\,{ y_4}}\over{3\,\sqrt{2}}}+{{2\,{ x_1}\,
 { y_3}}\over{3\,\sqrt{2}}}+{{2\,{ x_2}\,{ y_2}}\over{3}}+{{2
 \,{ x_3}\,{ y_1}}\over{\sqrt{30}}} \right) $$

$\bf6\otimes6=(1)\oplus(3)\oplus(5)\oplus(3'\oplus4)
\oplus(4\oplus5)\oplus(3\oplus3'\oplus5)$ 
$$ {\bf 6} \otimes {\bf 6} : {\bf 1} $$ 
$$\left( { x_1}\,{ y_6}-{ x_2}\,{ y_5}+{ x_3}\,
 { y_4}-{ x_4}\,{ y_3}+{ x_5}\,{ y_2}-{ x_6}\,
 { y_1} \right) $$

$$ {\bf 6} \otimes {\bf 6} : {\bf 3} $$ 
$$\left( {{{ x_1}\,{ y_5}}\over{\sqrt{5}}}-{{2\,\sqrt{2}\,
 { x_2}\,{ y_4}}\over{5}}+{{3\,{ x_3}\,{ y_3}}\over{5}}-
 {{2\,\sqrt{2}\,{ x_4}\,{ y_2}}\over{5}}+{{{ x_5}\,{ y_1}
 }\over{\sqrt{5}}} , {{{ x_1}\,{ y_6}}\over{\sqrt{2}}}-{{3\,
 { x_2}\,{ y_5}}\over{5\,\sqrt{2}}}+{{{ x_3}\,{ y_4}
 }\over{5\,\sqrt{2}}}+{{{ x_4}\,{ y_3}}\over{5\,\sqrt{2}}}-{{3
 \,{ x_5}\,{ y_2}}\over{5\,\sqrt{2}}}+{{{ x_6}\,{ y_1}
 }\over{\sqrt{2}}} , 
   \right.
 $$
 $$
 \left.
 {{{ x_2}\,{ y_6}}\over{\sqrt{5}}}-{{2\,
 \sqrt{2}\,{ x_3}\,{ y_5}}\over{5}}+{{3\,{ x_4}\,{ y_4}
 }\over{5}}-{{2\,\sqrt{2}\,{ x_5}\,{ y_3}}\over{5}}+{{{ x_6}
 \,{ y_2}}\over{\sqrt{5}}} \right) $$

$$ {\bf 6} \otimes {\bf 6} : {\bf 5} $$ 
$$\left( {{5\,{ x_1}\,{ y_4}}\over{\sqrt{10}}}-{{3\,{ x_2}\,
 { y_3}}\over{\sqrt{2}}}+{{3\,{ x_3}\,{ y_2}}\over{\sqrt{2}
 }}-{{5\,{ x_4}\,{ y_1}}\over{\sqrt{10}}} , {{5\,{ x_1}\,
 { y_5}}\over{\sqrt{5}}}-{{2\,{ x_2}\,{ y_4}}\over{\sqrt{2}
 }}+{{2\,{ x_4}\,{ y_2}}\over{\sqrt{2}}}-{{5\,{ x_5}\,
 { y_1}}\over{\sqrt{5}}} , 
   \right.
 $$
 $$
 \left.
 {{5\,{ x_1}\,{ y_6}}\over{\sqrt{6
 }}}+{{{ x_2}\,{ y_5}}\over{\sqrt{6}}}-{{4\,{ x_3}\,
 { y_4}}\over{\sqrt{6}}}+{{4\,{ x_4}\,{ y_3}}\over{\sqrt{6}
 }}-{{{ x_5}\,{ y_2}}\over{\sqrt{6}}}-{{5\,{ x_6}\,{ y_1}
 }\over{\sqrt{6}}} , 
   \right.
 $$
 $$
 \left.
 {{5\,{ x_2}\,{ y_6}}\over{\sqrt{5}}}-{{2\,
 { x_3}\,{ y_5}}\over{\sqrt{2}}}+{{2\,{ x_5}\,{ y_3}
 }\over{\sqrt{2}}}-{{5\,{ x_6}\,{ y_2}}\over{\sqrt{5}}} , {{5\,
 { x_3}\,{ y_6}}\over{\sqrt{10}}}-{{3\,{ x_4}\,{ y_5}
 }\over{\sqrt{2}}}+{{3\,{ x_5}\,{ y_4}}\over{\sqrt{2}}}-{{5\,
 { x_6}\,{ y_3}}\over{\sqrt{10}}} \right) $$

$$ {\bf 6} \otimes {\bf 6} : {\bf 3'} $$ 
$$\left( {{2\,{ x_4}\,{ y_6}}\over{\sqrt{5}}}-{{4\,\sqrt{2}\,
 { x_5}\,{ y_5}}\over{5}}+{{2\,{ x_6}\,{ y_4}}\over{
 \sqrt{5}}}-{{3\,{ x_1}\,{ y_4}}\over{\sqrt{5}}}+{{3\,{ x_2}
 \,{ y_3}}\over{5}}+{{3\,{ x_3}\,{ y_2}}\over{5}}-{{3\,
 { x_4}\,{ y_1}}\over{\sqrt{5}}} , 
   \right.
 $$
 $$
 \left.
 { x_1}\,{ y_6}+{{7\,
 { x_2}\,{ y_5}}\over{5}}-{{4\,{ x_3}\,{ y_4}}\over{5}}-
 {{4\,{ x_4}\,{ y_3}}\over{5}}+{{7\,{ x_5}\,{ y_2}}\over{5
 }}+{ x_6}\,{ y_1} ,
   \right.
 $$
 $$
 \left.
  {{3\,{ x_3}\,{ y_6}}\over{\sqrt{5}}}
 -{{3\,{ x_4}\,{ y_5}}\over{5}}-{{3\,{ x_5}\,{ y_4}
 }\over{5}}+{{3\,{ x_6}\,{ y_3}}\over{\sqrt{5}}}+{{2\,{ x_1}
 \,{ y_3}}\over{\sqrt{5}}}-{{4\,\sqrt{2}\,{ x_2}\,{ y_2}
 }\over{5}}+{{2\,{ x_3}\,{ y_1}}\over{\sqrt{5}}} \right) $$

$$ {\bf 6} \otimes {\bf 6} : {\bf 4} $$ 
$$\left( {{{ x_2}\,{ y_6}}\over{\sqrt{5}}}+{{{ x_3}\,
 { y_5}}\over{5\,\sqrt{2}}}-{{2\,{ x_4}\,{ y_4}}\over{5}}+{{
 { x_5}\,{ y_3}}\over{5\,\sqrt{2}}}+{{{ x_6}\,{ y_2}
 }\over{\sqrt{5}}} , -{{{ x_4}\,{ y_6}}\over{\sqrt{10}}}+{{2\,
 { x_5}\,{ y_5}}\over{5}}-{{{ x_6}\,{ y_4}}\over{\sqrt{10
 }}}-{{{ x_1}\,{ y_4}}\over{\sqrt{10}}}+{{{ x_2}\,{ y_3}
 }\over{5\,\sqrt{2}}}+{{{ x_3}\,{ y_2}}\over{5\,\sqrt{2}}}-{{
 { x_4}\,{ y_1}}\over{\sqrt{10}}} , 
   \right.
 $$
 $$
 \left.
 -{{{ x_3}\,{ y_6}
 }\over{\sqrt{10}}}+{{{ x_4}\,{ y_5}}\over{5\,\sqrt{2}}}+{{
 { x_5}\,{ y_4}}\over{5\,\sqrt{2}}}-{{{ x_6}\,{ y_3}
 }\over{\sqrt{10}}}+{{{ x_1}\,{ y_3}}\over{\sqrt{10}}}-{{2\,
 { x_2}\,{ y_2}}\over{5}}+{{{ x_3}\,{ y_1}}\over{\sqrt{10
 }}} , {{{ x_1}\,{ y_5}}\over{\sqrt{5}}}+{{{ x_2}\,{ y_4}
 }\over{5\,\sqrt{2}}}-{{2\,{ x_3}\,{ y_3}}\over{5}}+{{{ x_4}
 \,{ y_2}}\over{5\,\sqrt{2}}}+{{{ x_5}\,{ y_1}}\over{\sqrt{5
 }}} \right) $$

$$ {\bf 6} \otimes {\bf 6} : {\bf 4} $$ 
$$\left( -{{\sqrt{2}\,{ x_2}\,{ y_6}}\over{5\,\sqrt{3}}}-{{
 { x_3}\,{ y_5}}\over{\sqrt{15}}}+{{{ x_5}\,{ y_3}}\over{
 \sqrt{15}}}+{{\sqrt{2}\,{ x_6}\,{ y_2}}\over{5\,\sqrt{3}}}+{{2
 \,\sqrt{2}\,{ x_1}\,{ y_2}}\over{5\,\sqrt{3}}}-{{2\,\sqrt{2}\,
 { x_2}\,{ y_1}}\over{5\,\sqrt{3}}} , 
   \right.
 $$
 $$
 \left.
 {{{ x_4}\,{ y_6}
 }\over{5\,\sqrt{3}}}-{{{ x_6}\,{ y_4}}\over{5\,\sqrt{3}}}+{{
 \sqrt{3}\,{ x_1}\,{ y_4}}\over{5}}+{{{ x_2}\,{ y_3}
 }\over{\sqrt{15}}}-{{{ x_3}\,{ y_2}}\over{\sqrt{15}}}-{{\sqrt{3
 }\,{ x_4}\,{ y_1}}\over{5}} , 
   \right.
 $$
 $$
 \left.
 -{{\sqrt{3}\,{ x_3}\,
 { y_6}}\over{5}}-{{{ x_4}\,{ y_5}}\over{\sqrt{15}}}+{{
 { x_5}\,{ y_4}}\over{\sqrt{15}}}+{{\sqrt{3}\,{ x_6}\,
 { y_3}}\over{5}}+{{{ x_1}\,{ y_3}}\over{5\,\sqrt{3}}}-{{
 { x_3}\,{ y_1}}\over{5\,\sqrt{3}}} , 
   \right.
 $$
 $$
 \left.
 {{2\,\sqrt{2}\,{ x_5}
 \,{ y_6}}\over{5\,\sqrt{3}}}-{{2\,\sqrt{2}\,{ x_6}\,{ y_5}
 }\over{5\,\sqrt{3}}}+{{\sqrt{2}\,{ x_1}\,{ y_5}}\over{5\,
 \sqrt{3}}}+{{{ x_2}\,{ y_4}}\over{\sqrt{15}}}-{{{ x_4}\,
 { y_2}}\over{\sqrt{15}}}-{{\sqrt{2}\,{ x_5}\,{ y_1}}\over{5
 \,\sqrt{3}}} \right) $$

$$ {\bf 6} \otimes {\bf 6} : {\bf 5} $$ 
$$\left( -{{14\,{ x_4}\,{ y_6}}\over{\sqrt{15}}}+{{14\,
 { x_6}\,{ y_4}}\over{\sqrt{15}}}+{{3\,{ x_1}\,{ y_4}
 }\over{\sqrt{15}}}+{{{ x_2}\,{ y_3}}\over{\sqrt{3}}}-{{
 { x_3}\,{ y_2}}\over{\sqrt{3}}}-{{3\,{ x_4}\,{ y_1}
 }\over{\sqrt{15}}} , 
   \right.
 $$
 $$
 \left.
 {{14\,{ x_5}\,{ y_6}}\over{\sqrt{30}}}-{{14
 \,{ x_6}\,{ y_5}}\over{\sqrt{30}}}-{{8\,{ x_1}\,{ y_5}
 }\over{\sqrt{30}}}-{{4\,{ x_2}\,{ y_4}}\over{\sqrt{3}}}+{{4\,
 { x_4}\,{ y_2}}\over{\sqrt{3}}}+{{8\,{ x_5}\,{ y_1}
 }\over{\sqrt{30}}} , 
   \right.
 $$
 $$
 \left.
 { x_1}\,{ y_6}+3\,{ x_2}\,{ y_5}+2
 \,{ x_3}\,{ y_4}-2\,{ x_4}\,{ y_3}-3\,{ x_5}\,
 { y_2}-{ x_6}\,{ y_1} , 
   \right.
 $$
 $$
 \left.
 -{{8\,{ x_2}\,{ y_6}}\over{
 \sqrt{30}}}-{{4\,{ x_3}\,{ y_5}}\over{\sqrt{3}}}+{{4\,
 { x_5}\,{ y_3}}\over{\sqrt{3}}}+{{8\,{ x_6}\,{ y_2}
 }\over{\sqrt{30}}}-{{14\,{ x_1}\,{ y_2}}\over{\sqrt{30}}}+{{14
 \,{ x_2}\,{ y_1}}\over{\sqrt{30}}} , 
   \right.
 $$
 $$
 \left.
 {{3\,{ x_3}\,{ y_6}
 }\over{\sqrt{15}}}+{{{ x_4}\,{ y_5}}\over{\sqrt{3}}}-{{
 { x_5}\,{ y_4}}\over{\sqrt{3}}}-{{3\,{ x_6}\,{ y_3}
 }\over{\sqrt{15}}}+{{14\,{ x_1}\,{ y_3}}\over{\sqrt{15}}}-{{14
 \,{ x_3}\,{ y_1}}\over{\sqrt{15}}} \right) $$
 
$$ {\bf 6} \otimes {\bf 6} : {\bf 5} $$ 
$$\left( -{{2\,{ x_4}\,{ y_6}}\over{\sqrt{15}}}-{{2\,{ x_5}
 \,{ y_5}}\over{\sqrt{6}}}-{{2\,{ x_6}\,{ y_4}}\over{\sqrt{15
 }}}+{{{ x_1}\,{ y_4}}\over{\sqrt{15}}}+{{{ x_2}\,{ y_3}
 }\over{\sqrt{3}}}+{{{ x_3}\,{ y_2}}\over{\sqrt{3}}}+{{
 { x_4}\,{ y_1}}\over{\sqrt{15}}} ,
   \right.
 $$
 $$
 \left.
  -{{3\,{ x_5}\,{ y_6}
 }\over{\sqrt{30}}}-{{3\,{ x_6}\,{ y_5}}\over{\sqrt{30}}}-{{
 { x_1}\,{ y_5}}\over{\sqrt{30}}}-{{{ x_2}\,{ y_4}}\over{
 \sqrt{3}}}-{{2\,{ x_3}\,{ y_3}}\over{\sqrt{6}}}-{{{ x_4}\,
 { y_2}}\over{\sqrt{3}}}-{{{ x_5}\,{ y_1}}\over{\sqrt{30}}}
  ,\ { x_6}\,{ y_6}+{ x_1}\,{ y_1} , 
    \right.
 $$
 $$
 \left.
  {{{ x_2}\,
 { y_6}}\over{\sqrt{30}}}+{{{ x_3}\,{ y_5}}\over{\sqrt{3}}}+
 {{2\,{ x_4}\,{ y_4}}\over{\sqrt{6}}}+{{{ x_5}\,{ y_3}
 }\over{\sqrt{3}}}+{{{ x_6}\,{ y_2}}\over{\sqrt{30}}}-{{3\,
 { x_1}\,{ y_2}}\over{\sqrt{30}}}-{{3\,{ x_2}\,{ y_1}
 }\over{\sqrt{30}}} ,
   \right.
 $$
 $$
 \left.
  -{{{ x_3}\,{ y_6}}\over{\sqrt{15}}}-{{
 { x_4}\,{ y_5}}\over{\sqrt{3}}}-{{{ x_5}\,{ y_4}}\over{
 \sqrt{3}}}-{{{ x_6}\,{ y_3}}\over{\sqrt{15}}}-{{2\,{ x_1}\,
 { y_3}}\over{\sqrt{15}}}-{{2\,{ x_2}\,{ y_2}}\over{\sqrt{6}
 }}-{{2\,{ x_3}\,{ y_1}}\over{\sqrt{15}}} \right) $$

$$ {\bf 6} \otimes {\bf 6} : {\bf 3} $$ 
$$\left( {{7\,\sqrt{5}\,{ x_5}\,{ y_6}}\over{\sqrt{2}}}+{{7\,
 \sqrt{5}\,{ x_6}\,{ y_5}}\over{\sqrt{2}}}-{{5\,{ x_1}\,
 { y_5}}\over{\sqrt{10}}}-5\,{ x_2}\,{ y_4}-5\,\sqrt{2}\,
 { x_3}\,{ y_3}-5\,{ x_4}\,{ y_2}-{{5\,{ x_5}\,
 { y_1}}\over{\sqrt{10}}} , 
   \right.
 $$
 $$
 \left.
 -7\,{ x_6}\,{ y_6}+{ x_1}\,
 { y_6}+5\,{ x_2}\,{ y_5}+10\,{ x_3}\,{ y_4}+10\,
 { x_4}\,{ y_3}+5\,{ x_5}\,{ y_2}+{ x_6}\,{ y_1}+7
 \,{ x_1}\,{ y_1} , 
   \right.
 $$
 $$
 \left.
 -{{5\,{ x_2}\,{ y_6}}\over{\sqrt{10}
 }}-5\,{ x_3}\,{ y_5}-5\,\sqrt{2}\,{ x_4}\,{ y_4}-5\,
 { x_5}\,{ y_3}-{{5\,{ x_6}\,{ y_2}}\over{\sqrt{10}}}-{{7
 \,\sqrt{5}\,{ x_1}\,{ y_2}}\over{\sqrt{2}}}-{{7\,\sqrt{5}\,
 { x_2}\,{ y_1}}\over{\sqrt{2}}} \right) $$

$$ {\bf 6} \otimes {\bf 6} : {\bf 3'} $$ 
$$\left(-4\,\sqrt{5}\,{ x_4}\,{ y_6}-10\,\sqrt{2}\,{ x_5}\,
 { y_5}-4\,\sqrt{5}\,{ x_6}\,{ y_4}-3\,\sqrt{5}\,{ x_1}\,
 { y_4}-15\,{ x_2}\,{ y_3}-15\,{ x_3}\,{ y_2}-3\,
 \sqrt{5}\,{ x_4}\,{ y_1} , 
   \right.
 $$
 $$
 \left.
 -18\,{ x_6}\,{ y_6}-{ x_1}
 \,{ y_6}-5\,{ x_2}\,{ y_5}-10\,{ x_3}\,{ y_4}-10\,
 { x_4}\,{ y_3}-5\,{ x_5}\,{ y_2}-{ x_6}\,{ y_1}+18
 \,{ x_1}\,{ y_1} , 
   \right.
 $$
 $$
 \left.
 3\,\sqrt{5}\,{ x_3}\,{ y_6}+15\,
 { x_4}\,{ y_5}+15\,{ x_5}\,{ y_4}+3\,\sqrt{5}\,{ x_6}
 \,{ y_3}-4\,\sqrt{5}\,{ x_1}\,{ y_3}-10\,\sqrt{2}\,
 { x_2}\,{ y_2}-4\,\sqrt{5}\,{ x_3}\,{ y_1} \right) $$

\end{document}